\documentclass[aps,pre,reprint,groupedaddress,showpacs,showkeys]{revtex4-1}
\usepackage{amsmath, amssymb}
\usepackage{graphicx}
\usepackage{dcolumn}
\usepackage{enumerate}
\usepackage{bm}

\begin{document}
\title{Nonlinear transverse cascade and two-dimensional magnetohydrodynamic subcritical turbulence in plane shear flows}

\author{G. R. Mamatsashvili}
\email{george.mamatsashvili@tsu.ge}
\affiliation{Department of
Physics, Faculty of Exact and Natural Sciences, Tbilisi State
University, Tbilisi 0179, Georgia}

\author{D. Z. Gogichaishvili}
\affiliation{Department of Physics, The University of Texas at
Austin, Austin, Texas 78712, USA}

\author{G. D. Chagelishvili}
\affiliation{Institute of Geophysics, Tbilisi State University,
Tbilisi 0193, Georgia} \affiliation{Abastumani Astrophysical
Observatory, Ilia State University, Tbilisi 0162, Georgia}

\author{W. Horton}
\affiliation{Institute for Fusion Studies, The University of Texas
at Austin, Austin, Texas 78712, USA}

\date{\today}

\begin{abstract}
We find and investigate via numerical simulations self-sustained two-dimensional
turbulence in a magnetohydrodynamic flow with a maximally simple
configuration: plane, noninflectional (with a constant shear of
velocity) and threaded by a parallel uniform background magnetic
field. This flow is spectrally stable, so the turbulence is
subcritical by nature and hence it can be energetically supported
just by transient growth mechanism due to shear flow nonnormality.
This mechanism appears to be essentially anisotropic in spectral
(wavenumber) plane and operates mainly for spatial Fourier harmonics
with streamwise wavenumbers less than a ratio of flow shear to the
Alfv\'{e}n speed, $k_y < S/u_A$ (i.e., the Alfv\'{e}n frequency is lower
than the shear rate). We focused on the analysis of the
character of nonlinear processes and underlying self-sustaining
scheme of the turbulence, i.e., on the interplay between linear
transient growth and nonlinear processes, in spectral plane. Our
study, being concerned with a new type of the energy-injecting
process for turbulence -- the transient growth, represents an
alternative to the main trends of MHD turbulence research. We find
similarity of the nonlinear dynamics to the related dynamics in
hydrodynamic flows -- to the \emph{bypass} concept of subcritical
turbulence. The essence of the analyzed nonlinear MHD processes
appears to be a transverse redistribution of kinetic and magnetic spectral
energies in wavenumber plane [as occurs in the related hydrodynamic
flow, see Horton et al., Phys. Rev. E {\bf 81}, 066304 (2010)] and
differs fundamentally from the existing concepts of (anisotropic
direct and inverse) cascade processes in MHD shear flows.
\end{abstract}


\maketitle

\section{Introduction}

The problem of the onset and self-sustenance of turbulence in
spectrally stable nonuniform flows is a challenge to fluid dynamics
research. The efforts in this direction significantly increased in
the 1990s with the understanding and rigorous description of the
nonnormal nature of nonuniform, or shear flows (see e.g.,
Refs.~\cite{Reddy_etal93, Trefethen_etal93, Schmid_Henningson01,
Criminale_etal03, Schmid07}) and its direct consequences, such as
the possibility of finite-time, or transient growth of perturbations
in spectrally stable shear flows (e.g.,
Refs.~\cite{Gustavsson91,Farrell_Ioannou93, Reddy_Henningson93,
Farrell_Ioannou00}). Classical (direct and inverse) nonlinear
cascade processes, even if anisotropic, are in fact unable to
provide self-sustenance of perturbations (turbulence) when
transiently (non-exponentially) growing modes are present in the
flow. In the case of a specific shear flow, however, turbulence can
self-organize and be self-sustained through the subtle interplay of
the linear transient and nonlinear processes, where the flow shear
acts, through the Reynolds stress, to continuously supply the
turbulence with energy thanks to an essential constructive feedback
provided by the nonlinear processes
\cite{Gebhardt_Grossmann94,Henningson_Reddy94,Baggett_etal95,Grossmann00,
Chapman02, Eckhardt_etal07}.

The direct (nonlinear) cascade -- a central process in Kolmogorov's
phenomenology -- is a consequence of the existence of the so-called
inertial range in spectral (Fourier, or wavenumber) space, which is
free from the action of linear energy-exchange processes and, in
fact, occupied by nonlinear transfers. Kolmogorov's classical theory
of forced turbulence in hydrodynamics (HD) is the following: large
scale (long wavelength) perturbations imposed on the flow are
transferred by a direct nonlinear cascade, through the inertial
range, to short wavelengths and, ultimately, to the dissipation
region. So, the direct cascade, together with linear instability and
dissipative phenomena, constitute the well-known scheme of forced
turbulence in HD. However, in spectrally stable shear flows, where
transient growth of perturbations is the only possibility, the balance
of processes leading to the self-sustenance of turbulence should be
completely different. The shear-induced transient growth mainly
depends on the orientation (and, to a lesser degree, on the value) of the perturbation
wavevector: the spatial Fourier harmonics of perturbations (SFHs) having a
certain orientation of the wavevector with respect to the shear flow,
can draw flow energy and get amplified, whereas harmonics having
another orientation of the wavevector give energy back to the flow and
decay. In other words, the linear energy-exchange processes are
strongly anisotropic in wavenumber ${\bf k}$-space and occur over a
broad range of wavenumbers without leaving a free room (i.e.,
inertial range) for the action of nonlinear processes only. This
might render Kolmogorov's phenomenology inapplicable to spectrally
stable shear flows. A strong anisotropy of the linear processes in
shear flows, in turn, leads to anisotropy of nonlinear processes in
${\bf k}$-space. In this case, as revealed in
Ref.~\cite{Horton_etal10}, even in the simplest HD shear flow with
linear shear, the dominant nonlinear process turns out to be not
a direct, but a \emph{transverse cascade}, that is, a transverse (angular)
redistribution of perturbation harmonics over different quadrants of
wavenumber plane (e.g., from quadrants where $k_xk_y>0$ to quadrants
where $k_xk_y<0$ or vice versa). The interplay of this nonlinear
redistribution with linear phenomena (transient growth) becomes
intricate: it can provide either positive or negative feedback. In
the case of positive feedback, the nonlinearity repopulates
transiently growing modes and contributes to the self-sustenance of
perturbations. This combined action of anisotropic linear and
nonlinear processes can, in turn, give rise to an anisotropic energy
spectrum, which, in general, is expected to differ from the
Kolmogorovian. As a result, the transverse cascade may naturally
appear to be a possible keystone of the \emph{bypass} concept of
subcritical turbulence in spectrally stable HD shear flows, which is
being actively discussed among the hydrodynamical community (see
e.g., Refs.~\cite{Grossmann00,
Chapman02,Rempfer03,Eckhardt_etal07}).

In this paper, we extend the above study of nonlinear processes in
HD flows to magnetohydrodynamic (MHD) flows and investigate
subcritical turbulence in the simplest, spectrally stable shear flow
of magnetized plasma. We present the results of direct numerical
simulations (DNS) in Fourier plane, demonstrating the dominance of the
transverse cascade in MHD shear flows too. Specifically, we consider
the dynamics of two-dimensional (2D, with zero spanwise wavenumber,
$k_z=0$) perturbations in unbounded incompressible MHD fluid flow
with linear shear of velocity threaded by a uniform background
magnetic field directed parallel to the flow. This flow
configuration is spectrally stable in the linear regime
\cite{Stern63,Ogilvie_Pringle96} and therefore should be dominated
by the above-mentioned shear-induced transient phenomena
\cite{Chagelishvili_etal97}. Our main goals are:
\begin{enumerate}[(i)]
\item
to examine subcritical transition to turbulence and subsequent
self-sustaining dynamics by DNS,
\item
to describe the general behavior of nonlinear processes (transfers) --
transverse cascade -- in the presence of shear by carrying out an
analysis of these processes in Fourier plane,
\item
to show that the nonlinear transverse cascade is a keystone of
self-sustaining dynamics of the turbulence in this simple open MHD
flow system.
\end{enumerate}
The last point will allow us to find out in what form the
\emph{bypass} concept of subcritical turbulence can be realized in
spectrally stable MHD shear flows.

MHD turbulence phenomenon is ubiquitous in nature and is very
important in engineering applications. So, it is natural that there
is an enormous amount of research devoted  to it, starting with
seminal papers \cite{Iroshnikov63} and \cite{Kraichnan65} and their
extensions \cite{Goldreich_Sridhar95,Boldyrev05}. To date, the main
trends, including cases of forced and freely decaying MHD turbulence
as well as MHD turbulence with a
background magnetic field, established over decades
have been thoroughly analyzed in a number of review articles and books
(see e.g., Refs.~\cite{Biskamp03,Mininni11,Brandenburg_Lazarian13}
and references therein). Most of these analyses commonly focus
on turbulence dynamics in wavenumber space. However, the case of MHD
turbulence in smooth shear flows that we study here involves fundamental
novelties: an energy-supplying process for turbulence is the flow
nonnormality induced linear transient growth. The latter
anisotropically injects energy into turbulence over a broad range of
lengthscales and, consequently, rules out the inertial range of the sole
activity of nonlinearity and leads to a complex interplay of linear
and nonlinear processes. These circumstances give rise to new type
of processes in turbulence dynamics that are not accounted for
in the main trends of MHD turbulence research.

Magnetized shear flows have been considered in a number of papers
\cite{Kim06, Douglas_etal08, Newton_Kim09}. However, the range of
target parameters adopted in these studies excludes transient growth
effects due to shear and novelties associated with it. So, these
investigations still belong to the existing trends of MHD turbulence
research. For instance, these studies consider the limit of a strong
background magnetic field, ${\bf B_0}$, along the flow, where the
Alfv\'{e}n frequency of modes with wavenumber ${\bf k}$,
$\omega_A={\bf k}\cdot{\bf B}_0/(4\pi \rho_0)^{1/2}$ ($\rho_0$ is
the equilibrium density), is larger than shear rate of the mean flow
and since transient phenomena responsible for energy injection from
shear flow into perturbation harmonics are inefficient in this case,
external forcing (peaked at certain wavenumbers) is included to
drive turbulence. In contrast to this, in our case, the magnetic
field is weak and the adopted parameters permit an effective
transient exchange of energy between the mean flow and the perturbation
harmonics; this actually should serve to drive turbulence
without any external forcing. In this regard, in
Refs.~\cite{Hawley_etal95,Fromang_Papaloizou07,Guan_etal09,Simon_Hawley09,Lesur_Longaretti11},
the dynamics of MHD turbulence is investigated in a somewhat similar
setup -- astrophysical (protoplanetary) disk flows with Keplerian
shear and an imposed large-scale magnetic field which is typically weak
(i.e., usual plasma $\beta \gg 1$ in disks, see e.g.
Ref.~\cite{Armitage11}). This means that there exists harmonics
whose Alfv\'{e}n frequency is smaller than shear parameter, as in
our case. However, in
Refs.~\cite{Fromang_Papaloizou07,Lesur_Longaretti11}, although
turbulence dynamics is analyzed in Fourier space, the magnetic field
is directed perpendicular to the flow and consequently shear-induced
transient phenomena differ from those studied here. On the other
hand, Refs.~\cite{Hawley_etal95,Guan_etal09,Simon_Hawley09}
similarly to our study, consider orientation for the magnetic field along
the mean flow (i.e., azimuthal for disk flows). They observe
three-dimensional (3D) self-sustained turbulence, which is expected
to be governed by transient processes of a type similar to those of the 2D
shear turbulence studied here, but since the turbulence dynamics
(energy injection and transfers) was not investigated in
spectral space in those studies, identification of shear-induced effects is not
straightforward in their analysis.

The Earth's magnetosphere, created by the interaction of the solar
wind with the Earth's magnetic field, represents a huge
``laboratory'' of various MHD turbulence. In different parts of this
laboratory (e.g., ion foreshock, magnetosheath, LL magnetopause,
polar cusps, ionosphere, magnetotail) characteristic parameters
vary greatly from each other. There are shear flows, different
orientations of the magnetic field, different values of the plasma
$\beta$ parameter, anisotropic magnetic pressure, magnetic
reconnection, etc. (see e.g., Ref.~\cite{Zimbardo_etal10} for a
recent review). Evidently, it is hard to seek an immediate
realization of the proposed scheme of MHD shear turbulence in the
magnetized environment of the Earth. Still, certain areas can be
identified where a similar configuration and course of events are
realized. This, first of all, implies high-$\beta$ regions with
shear flows and a mean magnetic field parallel to the flow velocity.
Generally, such regions are in the magnetotail, magnetosheath and
cusp, but a definite view can be obtained after a detailed
investigation of the dynamical processes therein.

The specific nature of nonlinear processes, which we will focus on in our
study is, in many respects, a consequence of the shear-induced transient
linear dynamics described in
Refs.~\cite{Balbus_Hawley92,Chagelishvili_etal97,Dimitrov_etal11,Gogichaishvili_etal13}.
We particularly follow a recent paper \cite{Gogichaishvili_etal13},
where the linear dynamics of pseudo-Alfv\'{e}n waves (P-AWs) and
shear-Alfv\'{e}n waves (S-AWs) is described in a 3D MHD flow with
linear shear and parallel magnetic field. Specifically, it is shown
there that:
\begin{enumerate}
\item
Counter-propagating P-AWs are coupled to each other, while S-AWs
are not coupled with each other, but are asymmetrically coupled to
P-AWs; S-AWs do not participate in the linear dynamics of P-AWs,
\item
The linear coupling of counter-propagating waves determines the
transient growth (overreflection).
\item
The transient growth of S-AWs is somewhat smaller compared with that
of P-AWs,
\item
Waves with a smaller streamwise wavenumber, $k_y$, exhibit
stronger transient growth,
\item
Maximal transient growth (and overreflection) of the wave energy
occurs for 2D waves with $k_z=0$.
\end{enumerate}
These preliminary linear results served as a natural starting point
of the present study of nonlinear dynamics of 2D perturbations with
$k_z=0$ and white-noise initial spectrum in ${\bf k}$-plane using
DNS with a spectral code.

The paper is organized as follows. Sec. II is devoted to the
physical model and derivation of dynamical equations in spectral
plane. The DNS of the turbulence dynamics is presented in Sec.
III. In Sec. IV, we perform analysis of the numerical results
focusing on the activity of linear and nonlinear processes in
spectral plane. A summary and discussion are given in Section V.

\section{Physical model and equations}

The motion of an incompressible conducting fluid with constant
viscosity, $\nu$, and Ohmic resistivity, $\eta$, is governed by the
basic equations of MHD
\begin{equation}
\frac{\partial {\bf U}}{\partial t}+\left({\bf U}\cdot \nabla
\right){\bf U}=-\frac{\nabla P}{\rho}+\frac{\left({\bf B}\cdot\nabla
\right){\bf B}}{4\pi \rho}+\nu\nabla^2 {\bf U},
\end{equation}
\begin{equation}
\frac{\partial {\bf B}}{\partial t}=\nabla\times \left( {\bf
U}\times {\bf B}\right)+\eta\nabla^2{\bf B},
\end{equation}
\begin{equation}
\nabla\cdot {\bf U}=0,
\end{equation}
\begin{equation}
\nabla\cdot {\bf B}=0,
\end{equation}
where $\rho$ is the fluid density, ${\bf U}$ is the velocity, ${\bf
B}$ is the magnetic field and $P$ is the total pressure equal to the
sum of the thermal and magnetic pressures.

Equations (1)-(4) have a stationary equilibrium solution -- an
unbounded plane Couette flow along the $y-$axis with linear shear of
velocity in the the $x$-direction, ${\bf U}_0=(0,-Sx,0)$, and
threaded by a uniform background magnetic field parallel to the flow,
${\bf B}_0=(0,B_{0y},0)$. Without loss of generality, the
constant shear parameter $S$ and $B_{0y}$ are chosen to be positive,
$S, B_{0y}>0$. The equilibrium density $\rho_0$ and total pressure $P_0$
are spatially constant. Such a simple configuration of an unbounded
flow with a linear shear of the velocity profile corresponds, for
example, to plasma flow in astrophysical accretion disks in the
framework of the widely used local shearing box approximation (e.g.,
Ref.~\cite{Hawley_etal95}) as well as to flows of magnetized plasma
in the laboratory (e.g., Refs.~\cite{Kim06, Douglas_etal08}). It allows
us to grasp key effects of shear on the perturbation dynamics and,
ultimately, on the resulting MHD turbulent state in kinematically
nonuniform plasma flows.

Consider 2D perturbations of the velocity, total pressure and
magnetic field, ${\bf u}, p$ and ${\bf b}$, which are independent of the
vertical $z$-coordinate ($\partial/\partial z=0$), about the
equilibrium. In this case, the evolution in the horizontal
$(x,y)-$plane is decoupled from that of the $z-$components of the
perturbed velocity and magnetic field, so we set them to zero,
$u_z=b_z=0$. Representing the total fields as the sum of the
equilibrium and perturbed values, ${\bf U}={\bf U}_0+{\bf u},
P=P_0+p$ and ${\bf B}={\bf B}_0+{\bf b}$, substituting these into Eqs.
(1)-(4) and rearranging the nonlinear terms with the help of Eqs.
(3) and (4), we arrive at the following system governing the
dynamics of perturbations with arbitrary amplitude
\begin{multline}\label{1}
\left(\frac{\partial}{\partial t} - Sx \frac{\partial}{\partial y}
\right)u_x = - \frac{1}{\rho_0}\frac{\partial p}{\partial
x}+\frac{B_{0y}}{4\pi\rho_0}\frac{\partial b_x}{\partial
y}+\nu\nabla^2u_x+\\+\frac{\partial}{\partial
y}\left(\frac{b_xb_y}{4\pi\rho_0}-u_xu_y\right)+
\frac{\partial}{\partial
x}\left(\frac{b_x^2}{4\pi\rho_0}-u_x^2\right),
\end{multline}
\begin{multline}\label{2}
\left(\frac{\partial}{\partial t} - Sx \frac{\partial}{\partial y}
\right)u_y = Su_x-\frac{1}{\rho_0}\frac{\partial p}{\partial
y}+\frac{B_{0y}}{4\pi\rho_0}\frac{\partial b_y}{\partial
y}+\nu\nabla^2u_y+\\+\frac{\partial}{\partial
x}\left(\frac{b_xb_y}{4\pi\rho_0}-u_xu_y\right)+\frac{\partial}{\partial
y}\left(\frac{b_y^2}{4\pi\rho_0}-u_y^2\right)
\end{multline}
\begin{multline}\label{3}
\left(\frac{\partial}{\partial t} - Sx \frac{\partial}{\partial y}
\right)b_x = B_{0y}\frac{\partial u_x}{\partial
y}+\eta\nabla^2b_x+\\+\frac{\partial}{\partial y}
\left(u_xb_y-u_yb_x \right),
\end{multline}
\begin{multline}\label{4}
\left(\frac{\partial}{\partial t} - Sx \frac{\partial}{\partial y}
\right)b_y = - Sb_x+B_{0y}\frac{\partial u_y}{\partial
y}+\eta\nabla^2b_y-\\-\frac{\partial}{\partial x}
\left(u_xb_y-u_yb_x \right),
\end{multline}
\begin{equation}\label{5}
\frac{\partial u_x}{\partial x}+\frac{\partial u_y}{\partial y}=0,
\end{equation}
\begin{equation}\label{6}
\frac{\partial b_x}{\partial x}+\frac{\partial b_y}{\partial y}=0.
\end{equation}
We solve Eqs. (5)-(10) in a rectangular 2D domain with sizes
$L_x$ and $L_y$, respectively, in the $x-$ and $y-$directions,
$-L_x/2\leq x \leq L_x/2$ and $-L_y/2\leq y \leq L_y/2$, divided into
$N_x\times N_y$ cells. Since we consider an unbounded flow with
linear shear, we adopt boundary conditions commonly used in similar
cases of MHD simulations of astrophysical disk flows in the local
shearing box approximation (e.g.,
Refs.~\cite{Hawley_etal95,Fromang_Papaloizou07,Guan_etal09,Lesur_Longaretti11,Davis_etal10}).
Namely, for the perturbations of all quantities, we impose
periodic boundary conditions in the $y-$direction and
shearing-periodic in the $x-$direction. That is, the $x-$boundaries
are initially periodic, but shear with respect to each other as time
goes by, becoming again periodic at discrete moments
$t_n=nL_y/SL_x$, where $n=1,2,...$ is a positive integer. This can
be written as
\[
f(x,y,t)=f(x+L_x,y-SL_xt,t)~~~~~~~(x~{\rm boundary}),
\]
\[
f(x,y,t)=f(x,y+L_y,t)~~~~~~~(y~{\rm boundary}),
\]
where $f\equiv ({\bf u},p,{\bf b})$ denotes any of the perturbed
quantities. These boundary conditions ensure natural evolution of
shearing plane waves within the domain, as it would be in an
unbounded constant shear flow.

\subsection{Energy equation}

In this subsection, we derive dynamical equations for kinetic and
magnetic energies in order to gain insight into the interplay of the
flow shear and nonlinearity in the self-sustenance of perturbations.
The perturbation kinetic and magnetic energies are defined,
respectively, as
\[
E_K=\frac{\rho_0{\bf u}^2}{2}, ~~~ E_M=\frac{{\bf b}^2}{8\pi}.
\]
Using the main Eqs. (5)-(10) and the above shearing box boundary
conditions, after some algebra, we can readily derive the evolution
equation for the domain-averaged kinetic and magnetic energies
\begin{multline}
\frac{d}{dt}\langle E_K \rangle=S\left\langle\rho_0
u_xu_y\right\rangle+\frac{B_{0y}}{4\pi}\left\langle
u_x\frac{\partial b_x}{\partial y}+u_y\frac{\partial b_y}{\partial
y}\right\rangle+\\+\frac{1}{4\pi}\left\langle u_xb_y\frac{\partial
b_x}{\partial y}+\frac{u_x}{2}\frac{\partial b_x^2}{\partial
x}+\frac{u_y}{2}\frac{\partial b_y^2}{\partial y}+
u_yb_x\frac{\partial b_y}{\partial x} \right\rangle - \\
-\rho_0\nu\langle \left(\nabla u_x\right)^2+\left(\nabla
u_y\right)^2 \rangle,
\end{multline}
\begin{multline}
\frac{d}{dt}\langle E_M \rangle=S\left\langle
-\frac{b_xb_y}{4\pi}\right\rangle+\frac{B_{0y}}{4\pi}\left\langle
b_x\frac{\partial u_x}{\partial y}+b_y\frac{\partial u_y}{\partial
y}\right\rangle+\\+\frac{1}{4\pi}\left\langle
b_x\frac{\partial}{\partial y}(u_xb_y)+\frac{b_x^2}{2}\frac{\partial
u_x}{\partial x}+\frac{b_y^2}{2}\frac{\partial u_y}{\partial y}+
b_y\frac{\partial}{\partial x}(u_yb_x)\right\rangle - \\
-\frac{\eta}{4\pi}\langle \left(\nabla b_x\right)^2+\left(\nabla
b_y\right)^2 \rangle,
\end{multline}
where the angle brackets denote a spatial average, $\langle ...
\rangle=\int\int ...~ dxdy/L_xL_y$, with the integral being taken
over an entire domain. Adding up Eqs. (11) and (12), the cross terms
of linear origin, proportional to $B_{0y}$, and nonlinear terms
cancel out due to the boundary conditions and we obtain the equation
for the total energy $E=E_K+E_M$,
\begin{multline}
\frac{d\langle E \rangle}{dt}=S\left\langle
\rho_0u_xu_y-\frac{b_xb_y}{4\pi}\right\rangle-\\-\rho_0\nu\langle
\left(\nabla u_x\right)^2+\left(\nabla u_y\right)^2
\rangle-\frac{\eta}{4\pi}\langle\left(\nabla
b_x\right)^2+\left(\nabla b_y\right)^2\rangle.
\end{multline}
The first term on the right hand side of Eq. (13) is the shear
parameter, $S$, multiplied by the total stress in the angle
brackets. The total stress is the sum of the Reynolds, $\rho_0u_xu_y$,
and Maxwell, $-b_xb_y/4\pi$, stresses which describe, respectively, the
exchange of kinetic and magnetic energies between perturbations and
the background flow in Eqs. (11) and (12). Note that they originate
from the linear terms proportional to shear on the right hand sides of Eqs. (6) and (8).
These stresses also determine the rate of momentum transport (see
e.g., Refs.~\cite{Hawley_etal95,Balbus03,Douglas_etal08}) and thus
are one of the important quantities characterizing shear flow
turbulence. The second and third terms describe energy dissipation
due to viscosity and resistivity, respectively. Note that the net
contribution from nonlinear terms has canceled out in the total
energy evolution Eq. (13) after averaging over the domain. Thus,
only Reynolds and Maxwell stresses can supply perturbations with
energy, extracting it from the mean flow due to shear; the other two
terms are negative definite and dissipative. In the case of shear
flow turbulence studied below, these stresses ensure energy
injection into turbulent fluctuations. The nonlinear terms, not
directly tapping into the shear flow energy and therefore not changing
the total perturbation energy, serve only to redistribute energy
gained by means of the stresses among Fourier harmonics of
perturbations with different wavenumbers (see below). In the absence
of shear ($S=0$), the contribution from the Reynolds and Maxwell
stresses disappears in Eq. (13) and hence the total perturbation
energy cannot grow, gradually decaying due to viscosity and
resistivity.

\subsection{Spectral representation of the equations}

Before proceeding further, we normalize the variables by taking the
shear time, $S^{-1}$, as the unit of time, the Alfv\'{e}n speed,
$u_A=B_{0y}/(4\pi \rho_0)^{1/2}$, as the unit of velocity,
$\ell\equiv u_AS^{-1}$ as the unit of length and $B_{0y}$ as the
unit of the magnetic field perturbations,
\[
St\rightarrow t,~~~\left(\frac{x}{\ell},\frac{y}{\ell} \right)
\rightarrow (x,y),~~~\frac{{\bf u}}{u_A}\rightarrow {\bf u},
\]
\[
\frac{p}{\rho_0u_A^2}\rightarrow p,~~~\frac{{\bf
b}}{B_{0y}}\rightarrow {\bf
b},~~~\frac{E_{K,M}}{\rho_0u_A^2}\rightarrow E_{K,M}.
\]
Viscosity and resistivity are characterized by hydrodynamic, ${\rm
Re}$, and magnetic, ${\rm Rm}$, Reynolds numbers defined here, for
convenience, in terms of $u_A$ and $\ell$ as
\[
{\rm Re}= \frac{u_A\ell}{\nu}=\frac{u_A^2}{\nu S},~~~~~{\rm
Rm}=\frac{u_A\ell}{\eta}= \frac{u_A^2}{\eta S}.
\]
These numbers are also referred to, respectively, as viscous and
resistive Elsasser numbers (e.g., Ref.~\cite{Lesur_Longaretti11}).
The strength of the imposed mean magnetic field is measured by the
ratio of the mean flow kinetic energy to the magnetic energy within
the domain
\[
\beta=\frac{\pi\rho_0S^2L_x^2}{3B_{0y}^2}=\frac{S^2L_x^2}{12u_A^2}=\frac{L_x^2}{12
\ell^2}.
\]

For further analysis, we need to do a spectral representation of the
main equations. We decompose the perturbations into spatial Fourier
harmonics (SFHs)
\begin{equation}
f({\bf r},t)=\int \bar{f}({\bf k},t)\exp\left({\rm i}{\bf
k}\cdot{\bf r} \right)d^2{\bf k}
\end{equation}
where, as before, $f\equiv ({\bf u},p,{\bf b})$ denotes the
perturbations and $\bar{f}\equiv (\bar{\bf u}, \bar{p}, \bar{\bf
b})$ is their corresponding Fourier transforms ($k_z=0$ for
$z-$independent 2D perturbations and $d^2{\bf k}\equiv dk_xdk_y$).
Substituting decomposition (14) into Eqs. (5)-(10) and taking into
account the above normalization, we arrive at the following
equations governing the dynamics of perturbation SFHs in spectral
plane
\begin{multline}
\left(\frac{\partial}{\partial t}+k_y\frac{\partial}{\partial
k_x}\right)\bar{u}_x=-{\rm i}k_x\bar{p}+{\rm
i}k_y\bar{b}_x-\frac{k^2}{\rm Re}\bar{u}_x+\\+{\rm i}k_y N_1+{\rm
i}k_xN_2,
\end{multline}
\begin{multline}
\left(\frac{\partial}{\partial t}+k_y\frac{\partial}{\partial
k_x}\right)\bar{u}_y=\bar{u}_x-{\rm i}k_y\bar{p}+{\rm
i}k_y\bar{b}_y-\frac{k^2}{\rm Re}\bar{u}_y+\\+{\rm i}k_xN_1+{\rm
i}k_yN_3,
\end{multline}
\begin{equation}
\left(\frac{\partial}{\partial t}+k_y\frac{\partial}{\partial
k_x}\right)\bar{b}_x={\rm i}k_y\bar{u}_x-\frac{k^2}{\rm
Rm}\bar{b}_x+{\rm i}k_yN_4,
\end{equation}
\begin{equation}
\left(\frac{\partial}{\partial t}+k_y\frac{\partial}{\partial
k_x}\right)\bar{b}_y=-\bar{b}_x+{\rm i}k_y\bar{u}_y-\frac{k^2}{\rm
Rm}\bar{b}_y-{\rm i}k_xN_4
\end{equation}
\begin{equation}
k_x\bar{u}_x+k_y\bar{u}_y=0,
\end{equation}
\begin{equation}
k_x\bar{b}_x+k_y\bar{b}_y=0,
\end{equation}
where $k^2=k_x^2+k_y^2$ (wavenumbers are normalized by $\ell^{-1}$).
These spectral equations contain the linear as well as the
nonlinear, $N_1({\bf k},t)$, $N_2({\bf k},t)$, $N_3({\bf k},t)$ and
$N_4({\bf k},t)$, terms that are the Fourier transforms of
corresponding linear and nonlinear terms in the original Eqs.
(5)-(10). The latter are given by
\begin{multline*}
N_1({\bf k},t)=\\=\int d^2{\bf k'}\left[\bar{b}_x({\bf
k'},t)\bar{b}_y({\bf k}-{\bf k'},t)-\bar{u}_x({\bf
k'},t)\bar{u}_y({\bf k}-{\bf k'},t)\right]
\end{multline*}
\begin{multline*}
N_2({\bf k},t)=\\=\int d^2{\bf k'}\left[\bar{b}_x({\bf
k'},t)\bar{b}_x({\bf k}-{\bf k'},t)-\bar{u}_x({\bf
k'},t)\bar{u}_x({\bf k}-{\bf k'},t)\right]
\end{multline*}
\begin{multline*}
N_3({\bf k},t)=\\=\int d^2{\bf k'}\left[\bar{b}_y({\bf
k'},t)\bar{b}_y({\bf k}-{\bf k'},t)-\bar{u}_y({\bf
k'},t)\bar{u}_y({\bf k}-{\bf k'},t)\right]
\end{multline*}
\begin{multline*}
N_4({\bf k},t)=\\=\int d^2{\bf k'}\left[\bar{u}_x({\bf
k'},t)\bar{b}_y({\bf k}-{\bf k'},t)-\bar{u}_y({\bf
k'},t)\bar{b}_x({\bf k}-{\bf k'},t)\right]
\end{multline*}
and describe nonlinear triad interactions among velocity and
magnetic field components of SFHs with different wavenumbers in
Fourier ${\bf k}$-plane. Equations (15)-(20), which are the basis for
subsequent analysis, involve two free dissipative parameters ${\rm
Re}$ and ${\rm Rm}$. Since we consider a finite domain in physical
$(x,y)$-plane, the perturbation dynamics also depends on the
smallest wavenumber available in this domain, or equivalently on its
sizes $L_x$ and $L_y$, which are the other two free parameters of
the problem. Given these parameters and specific initial conditions,
Eqs. (15)-(20) fully determine the nonlinear dynamics of the
considered system in Fourier plane. These equations form the
mathematical basis of our main goal -- to investigate the character
of nonlinear processes and self-sustaining scheme of the
(subcritical) MHD turbulence in ${\bf k}$-plane in this constant
shear flow. Since energy spectra and nonlinear transfers relate to
energy equations, following
Refs.~\cite{Chagelishvili_etal02,Alexakis_etal07,Fromang_Papaloizou07,Simon_etal09,Horton_etal10,Lesur_Longaretti11},
below we derive equations governing the evolution of kinetic and
magnetic spectral energies.

Multiplying Eqs. (15) and (16), respectively, by $\bar{u}_x^{\ast}$
and $\bar{u}_y^{\ast}$, combining and adding its complex conjugate,
we arrive at the following equation for the nondimensional kinetic
spectral energy $\bar{E}_K=|\bar{u}_x|^2+|\bar{u}_y|^2$,
\begin{equation}
\frac{\partial \bar{E}_K}{\partial t}+\frac{\partial}{\partial
k_x}\left(k_y\bar{E}_K\right)=I_K+I_{K-M}+D_K+N_K,
\end{equation}
where
\[
I_K=\bar{u}_x\bar{u}_y^{\ast}+\bar{u}_x^{\ast}\bar{u}_y=-\frac{2k_xk_y}{k^2}\bar{E}_K,~~~D_K=-\frac{2k^2}{\rm
Re}\bar{E}_K,
\]
\[
I_{K-M}={\rm
i}k_y\left(\bar{u}_x^{\ast}\bar{b}_x+\bar{u}_y^{\ast}\bar{b}_y-\bar{u}_x\bar{b}_x^{\ast}-\bar{u}_y\bar{b}_y^{\ast}\right),
\]
and the nonlinear kinetic transfer function $N_K({\bf k},t)$ is
given by
\begin{multline*}
N_K({\bf k},t)={\rm
i}(k_y\bar{u}_x^{\ast}+k_x\bar{u}_y^{\ast})N_1({\bf k},t)+\\+{\rm
i}k_x\bar{u}_x^{\ast}[N_2({\bf k},t)-N_3({\bf k},t)]+{\rm c. c.}~.
\end{multline*}
Similarly, multiplying Eqs. (17) and (18), respectively, by
$\bar{b}_x^{\ast}$ and $\bar{b}_y^{\ast}$, combining and adding its
complex conjugate, we obtain the evolution equation for the
nondimensional magnetic spectral energy
$\bar{E}_M=|\bar{b}_x|^2+|\bar{b}_y|^2$,
\begin{equation}
\frac{\partial \bar{E}_M}{\partial t}+\frac{\partial}{\partial
k_x}\left(k_y\bar{E}_M\right)=I_M+I_{M-K}+D_M+N_M,
\end{equation}
where
\[
I_M=-\bar{b}_x\bar{b}_y^{\ast}-\bar{b}_x^{\ast}\bar{b}_y=\frac{2k_xk_y}{k^2}\bar{E}_M,
\]
\[
I_{M-K}=-I_{K-M},~~~D_M=-\frac{2k^2}{\rm Rm}\bar{E}_M
\]
and the nonlinear magnetic transfer function $N_M({\bf k},t)$ is
given by
\[
N_M({\bf k},t)={\rm
i}(k_y\bar{b}_x^{\ast}-k_x\bar{b}_y^{\ast})N_4({\bf k},t)+{\rm
c.c.}~.
\]
By inspection of Eqs. (21) and (22), one can distinguish five basic
processes underlying the dynamics of $\bar{E}_K$ and $\bar{E}_M$:
\begin{enumerate}
\item
The quantities $k_y\bar{E}_K$ and $k_y\bar{E}_M$ in the second terms
on the left hand sides of Eqs. (21) and (22) are, respectively, the
fluxes of the kinetic and magnetic spectral energies parallel to the
$k_x-$axis. These terms are of linear origin, coming from the
convective derivative on the left hand sides of the main Eqs.
(5)-(10) and therefore correspond to the advection by the mean flow.
In other words, background shear flow makes the spectral energies
(Fourier transforms) ``drift'' in ${\bf k}-$plane, and SFHs with $k_y>0$
and $k_y<0$ travel, respectively, along and opposite the $k_x-$axis
at a speed $|k_y|$, whereas SFHs with $k_y=0$ are not advected
by the flow. Since $\int d^2{\bf k}\partial
(k_y\bar{E}_{K,M})/\partial k_x=0$, this drift only transports SFHs
parallel to the $k_x-$axis, without changing the total kinetic and
magnetic energies.
\item
The first terms on the right hand sides, $I_K$ and $I_M$, are
associated with shear, i.e., they originate from linear terms
proportional to the shear parameter on the right hand side of Eqs. (6) and (8), and
describe energy exchange between the mean flow and individual SFHs.
These terms are related to the domain-averaged nondimensional
Reynolds and Maxwell stresses entering Eqs. (11) and (12) through
\[
\langle u_xu_y\rangle=\frac{1}{2}\int I_K({\bf k},t)d^2{\bf k},
\]
\[
\langle -b_xb_y\rangle=\frac{1}{2}\int I_M({\bf k},t)d^2{\bf k}
\]
and therefore serve as a main source of energy for SFHs (with
$k_y\neq 0$) at the expense of which they can undergo amplification.
This shear-induced growth of perturbation SFHs is in fact linear by
nature and has a transient character due to the drift in ${\bf
k}-$plane
\cite{Balbus_Hawley92,Chagelishvili_etal97,Dimitrov_etal11,Mamatsashvili_etal13,Pessah_Chan12}.
The SFHs, drifting parallel to the $k_x-$axis, go through
dynamically important regions in spectral plane, where
\emph{energy-supplying} linear terms, $I_K$ and $I_M$, and
\emph{redistributing} nonlinear terms, $N_K$ and $N_M$, are at work
from small and intermediate wavenumbers almost up to the dissipation
region at large wavenumbers (see e.g., Fig. 6). In the case of
turbulence studied below, $I_K$ and $I_M$ describe the injection,
respectively, of kinetic and magnetic energies into turbulent
fluctuations as a function of wavenumbers (see also
Refs.~\cite{Fromang_Papaloizou07,Lesur_Longaretti11}).
\item
The second, cross terms on the right hand sides, $I_{K-M}$ and
$I_{M-K}$, describe the exchange between kinetic and magnetic spectral
energies. They have opposite signs and therefore cancel out in the
total energy budget of SFHs [see Eq. (24) below]. These terms are
also of linear origin, corresponding to terms proportional to
$B_{0y}$ (linearized magnetic tension and electromotive forces) in
Eqs. (5)-(8).
\item
The third terms on the right hand sides, $D_K$ and $D_M$, describe
the dissipation of kinetic and magnetic energies due to viscosity and
resistivity, respectively. Comparing these dissipation terms with
the energy-supplying terms $I_K$ and $I_M$, we see that viscous and
resistive dissipation are important at large wavenumbers $k \gtrsim
k_D={\rm min}(\sqrt{\rm Re}, \sqrt{\rm Rm})$, where $k_D$
denotes the effective wavenumber for which dissipation effects start
to play a role.
\item
The fourth terms on the right hand sides, $N_K$ and $N_M$, describe
nonlinear transfers, respectively, of kinetic and magnetic energies
among SFHs with different wavenumbers in ${\bf k}-$plane. It follows
from the definition of $N_K$ and $N_M$ that their sum integrated
over an entire wavenumber plane is equal to zero,
\begin{equation}
\int [N_K({\bf k},t)+N_M({\bf k},t)]d^2{\bf k}=0,
\end{equation}
which is, in fact, a direct consequence of the vanishing of the
nonlinear terms in the total energy Eq. (13) in real plane. This implies
that the main effect of nonlinearity is only to redistribute
(scatter) energy drawn from the mean flow among kinetic and magnetic
components of perturbation SFHs with different wavenumbers, while
leaving the total (kinetic plus magnetic) spectral energy summed
over all wavenumbers unchanged. In general, nonlinear transfer
functions, $N_K$ and $N_M$, play a central role in MHD turbulence
theory -- they determine cascades of spectral energies in ${\bf
k}-$space, leading to the development of their specific spectra.
These transfer functions are one of the main focuses of the present
analysis. We aim to explore how they operate in the presence of
shear, adopting the approach of
Refs.~\cite{Chagelishvili_etal02,Horton_etal10}, which numerically
studied the nonlinear dynamics of 2D perturbations in an HD Couette
flow by performing a full 2D Fourier analysis of individual terms in
the evolution equation for spectral energy, thus allowing for
anisotropy of spectra and cascades. In particular, we show
below that like that in the HD shear flow, nonlinear transfers in
the quasi-steady MHD shear turbulence result in the redistribution
of spectral energy among wavevector angles in ${\bf k}-$plane, which
we refer to as a nonlinear transverse cascade, in contrast to
classical HD or MHD turbulence without background shear flow, where
energy cascade processes change only the wavevector magnitude, $k=|{\bf
k}|$, of SFHs (see e.g., Ref.~\cite{Biskamp03}).
\end{enumerate}

Combining Eqs. (21) and (22), we obtain the equation for the total
spectral energy $\bar{E}=\bar{E}_K+\bar{E}_M$,
\begin{equation}
\frac{\partial \bar{E}}{\partial t}+\frac{\partial}{\partial
k_x}\left(k_y\bar{E}\right)=I_K+I_M+D_K+D_M+N_K+N_M.
\end{equation}
As mentioned above, the linear cross terms responsible for kinetic
and magnetic energy exchange are absent in this equation. The net
effect of the nonlinear terms in the total spectral energy budget
over all wavenumbers is zero according to Eq. (23). Thus, as
follows from Eq. (24), the only source for the total perturbation
energy is the integral over an entire spectral plane $\int
(I_K+I_M)d^2{\bf k}$ that extracts energy from a vast reservoir of
shear flow and injects it into perturbations. Since the terms $I_K$
and $I_M$, as noted above, are of linear origin, the energy
extraction and perturbation growth mechanisms are essentially linear
by nature. The role of nonlinearity is to continually provide, or
regenerate those SFHs in ${\bf k}-$plane that are able to undergo
transient growth, drawing on the mean flow energy, and in this way feed
the nonlinear state over long times. This scenario of a
self-sustained state, based on a subtle cooperation between linear
and nonlinear processes, is a keystone of the bypass concept of
subcritical turbulence in spectrally stable shear flows
\cite{Gebhardt_Grossmann94,Baggett_etal95,Grossmann00,Chapman02,Rempfer03,Eckhardt_etal07}.

\section{Nonlinear evolution}

We now turn to an analysis of the nonlinear evolution of perturbations
employing modern numerical methods. The main emphasis is on the
spectral aspect of the dynamics using the mathematical formalism
outlined in the previous section. We start a fiducial run by
imposing solenoidal random noise perturbations of the velocity and
magnetic field with spatially uniform rms amplitudes $\langle {\bf
u}^2 \rangle^{1/2}=\langle {\bf b}^2 \rangle^{1/2}=0.84$ on top of
the equilibrium. The computational domain is a square of size
$L_x\times L_y=400\times 400$ and resolution $N_x\times
N_y=512\times 512$. The reason for taking a large domain is to
encompass wavenumbers as small as possible at which, as
shown below, the effective transient amplification of SFHs and most
of dynamical activity take place. The minimum and maximum
wavenumbers of the domain are $k_{x,min}=k_{y,min}=2\pi/L_x=0.016$
and $k_{x,max}=k_{y,max}=\pi N_x/L_x=4.02$. The viscous and
resistive Reynolds numbers are fixed to the values ${\rm Re}={\rm
Rm}=5$ (corresponding to magnetic Prandtl number of unity ${\rm
Pr}={\rm Rm}/{\rm Re}=1$), so that the dissipation wavenumber,
$k_D$, falls in this range, $k_{D}=\sqrt{\rm
Re}=2.24<k_{x,max}$.\footnote{The usual Reynolds numbers defined in
terms of the half domain size $L_x/2$ and the mean flow velocity at
the domain boundary, $U_{0,max}=SL_x/2$, ${\rm
Re}^{\ast}=L_xU_{0,max}/2\nu, {\rm Rm}^{\ast}=L_xU_{0,max}/2\eta$,
are related to the Reynolds numbers used here by ${\rm
Re}^{\ast}=L_x^2{\rm Re}/4, {\rm Rm}^{\ast}=L_x^2{\rm Rm}/4$.
So, for $L_x=400$ and ${\rm Re}={\rm Rm}=5$, these numbers are
actually quite large ${\rm Re}^{\ast}={\rm Rm}^{\ast}=2.0\times
10^{5}$}  Note also that for the domain size $L_x=400$ the above
defined parameter $\beta=L_x^2/12=1.33\times 10^{4}$ is quite large,
indicating that the background magnetic field energy is small
compared to the kinetic energy of the mean flow and therefore the
flow can be regarded as weakly magnetized.

\begin{figure}
\includegraphics[width=\columnwidth]{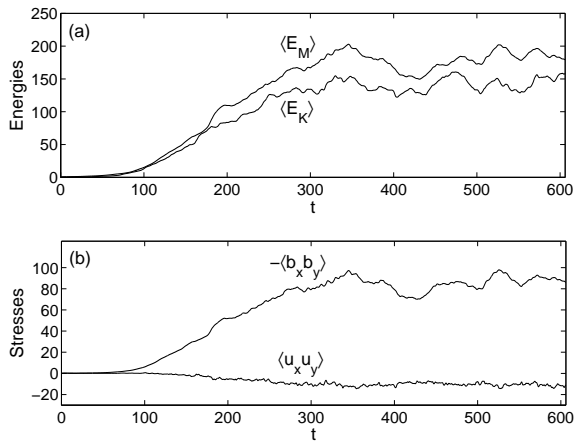}
\caption{Evolution of the domain-averaged (a) perturbed kinetic,
$\langle E_K \rangle$, and magnetic, $\langle E_M \rangle$, energies
as well as (b) the Reynolds and Maxwell stresses in the fiducial
run. Data have been boxcar-averaged over 60 shear times to make
the plot readable. In the beginning, all these quantities steadily
grow as a result of shear-induced transient amplification of
separate SFHs. Then, at about $t=250$, the amplification saturates
to a quasi-steady turbulent state that persists till the end of the
run. The magnetic energy is a bit higher than the kinetic one and
the positive Maxwell stress dominates over the negative Reynolds
stress.}
\end{figure}
\begin{figure*}
\includegraphics[width=\textwidth]{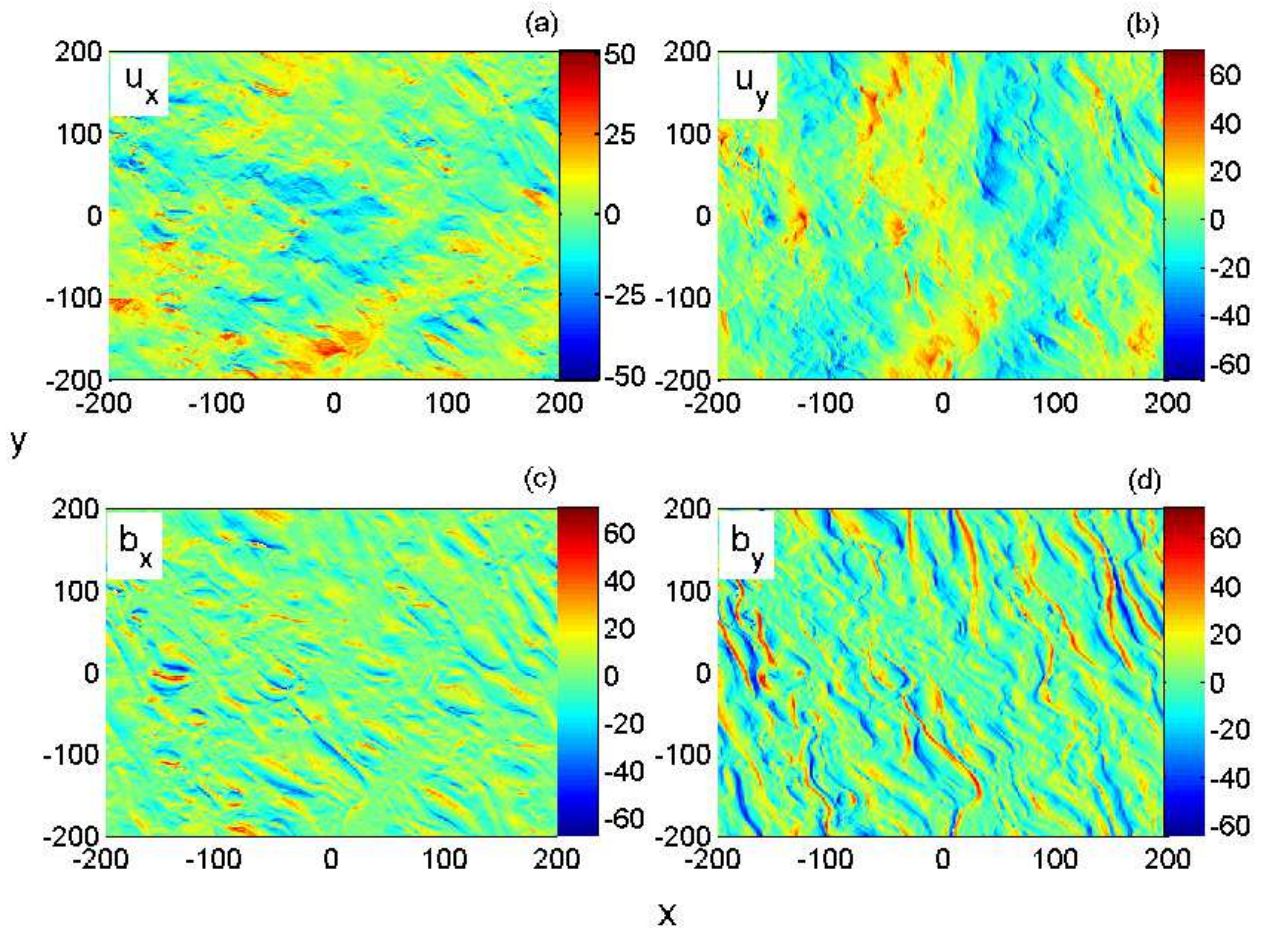}
\caption{(Color online) Distribution of [(a),(b)] the velocity and
[(c),(d)] the magnetic field components in $(x,y)-$plane in the fully
developed quasi-steady turbulence at $t=490$. This state is fairly
nonlinear: $u_x$ and $u_y$ vary within limits comparable to the
domain-averaged velocity of the background flow (in non-dimensional
units $\langle |U_0|\rangle=L_x/4=100$), while $b_x$ and $b_y$ are
much larger than the background magnetic field $B_{0y}=1$.
Structures in the $u_y$ and $b_y$ fields are elongated in the
$y-$direction due to shear.}
\end{figure*}

The subsequent time-evolution with these initial conditions was
followed to $t_f=600$ (i.e., for a total of 600 shear times) by
solving the basic Eqs. (5)-(10) using the spectral \textsc{snoopy} code
\footnote{The code is available for download at G. Lesur's web page
\texttt{http://ipag.obs.ujf-grenoble.fr/$\sim$lesurg/snoopy.html}}.
The mean magnetic field ${\bf B_0}$ is conserved with time, because
the domain-averaged fluctuating (turbulent) fields, as we checked,
remain zero, $\langle {\bf u}\rangle=\langle {\bf b}\rangle=0$,
during the whole run thanks to the shearing box boundary conditions.
The \textsc{snoopy} is a general purpose code, solving HD and MHD equations, including
shear, rotation, weak compressibility and several other physical
effects. It is based on a spectral (Fourier) method allowing for the
drift of harmonics in ${\bf k}$-space due to mean flow (i.e., the
shearing box boundary conditions are implemented in the code). The
Fourier transforms are computed using the FFTW 3 library. Nonlinear
terms are computed using a pseudo-spectral algorithm
\cite{Canuto_etal88} and antialiasing is enforced using the ``3/2''
rule. Time-integration is performed by a third order Runge-Kutta
scheme for nonlinear terms, whereas an implicit scheme is used for
viscous and resistive terms. This spectral scheme uses a periodic
remap algorithm in order to continually follow the smallest
wavenumber of the system in the sheared frame moving with the flow.
The code has been tested and extensively used in a number of fluid
dynamical and astrophysical contexts (see e.g.,
Refs.~\cite{Lesur_Longaretti11,Lesur_Longaretti07,Lesur_Ogilvie08,Lesur_Papaloizou10,
Longaretti_Lesur10,Rempel_etal10,Herault_etal11}).

Figure 1 shows the time-development of the domain-averaged perturbed
kinetic, $\langle E_K\rangle$, and magnetic, $\langle E_M \rangle$,
energies as well as the Reynolds, $\langle u_xu_y \rangle$, and
Maxwell $-\langle b_xb_y \rangle$ stresses. At the early stage of
evolution, they all increase as a result of linear transient growth
of separate SFHs contained in the initial conditions. Then, after
about 250 shear times, on reaching sufficient amplitudes in the
nonlinear regime, the energies and stresses settle down to a
quasi-steady state of sustained turbulence (see Fig. 2) that does
not decay and persists until the end of the simulation at $t_f=600$.
In this state, the kinetic and magnetic energies are comparable -- a
ratio of their domain- and time-averaged over the whole quasi-steady
state (denoted here and below, for the stresses, with double
brackets) values is $\langle\langle E_M
\rangle\rangle/\langle\langle E_K\rangle\rangle = 1.28$, that is,
there is a near equipartition of the energy between kinetic and
magnetic components. The Maxwell stress is much larger than the
Reynolds stress, indicating that the turbulent transport and energy
extraction from the mean flow are dominated by the magnetic field
perturbations. The average of the domain-averaged Maxwell stress over
the last 350 shear times is positive $\langle\langle
-b_xb_y\rangle\rangle=84.5$, while that of the domain-averaged
Reynolds stress is negative $\langle\langle
u_xu_y\rangle\rangle=-10.4$. As is seen from Eq. (13), the
domain-averaged total stress must necessarily be positive for
maintenance of turbulence and therefore it is the Maxwell stress
that plays a decisive role in this process -- counteracting
dissipation, it ensures continuous feeding and sustenance of the
turbulence at the expense of the mean shear flow.

The structure of the velocity and magnetic field in the quasi-steady
turbulent state (at $t=490$) is depicted in Fig. 2. These fields are
chaotic with $u_y$ and $b_y$ [Figs. 2(b) and 2(d)] having more
elongated features in the $y-$direction due to shear compared to
$u_x$ and $b_x$ [Figs. 2(a) and 2(c)]. At this time, the normalized
fluctuating velocity and magnetic field are comparable, $\langle
u_x^2\rangle=87.68$, $\langle u_y^2\rangle= 178.73$, $\langle
b_x^2\rangle=113.17, \langle b_y^2\rangle=238.64$ and are much
larger than their corresponding initial values. Also, the
$y-$components are larger than the $x$-ones: $\langle
u_x^2\rangle<\langle u_y^2\rangle$, $\langle b_x^2\rangle<\langle
b_y^2\rangle$, which holds throughout the run. Within the domain,
$u_x$ and $u_y$ reach maximum values $|u_x|_{\rm max}=51.53$ and
$|u_y|_{\rm max}= 70.15$ comparable to the average background flow
velocity, $\langle |U_0| \rangle=L_x/4=100$, and the $b_x$ and $b_y$
have grown much larger, $|b_x|_{\rm max}=70.61$ and $|b_y|_{\rm
max}=72.78$, than the mean field $B_{0y}=1$. So, this quasi-steady
MHD turbulence can be viewed as being strongly nonlinear and weakly
magnetized, since $\langle {\bf b}^2\rangle^{1/2}\gg B_{0y}$.

The general behavior of the domain-averaged kinetic and magnetic
energies and stresses with time obtained here in the 2D case is
qualitatively consistent with that typically found in similar, but
3D simulations of MHD turbulence driven by the magnetorotational
instability (MRI) in local models of accretion disks with a net
toroidal magnetic field along the disk flow
\cite{Hawley_etal95,Guan_etal09,Simon_Hawley09}, as in the present
setup. In both cases, there are no exponentially growing modes in
the considered unbounded constant shear flows in the classical sense
of linear stability analysis \cite{Stern63,Ogilvie_Pringle96}, i.e.,
the flows are spectrally stable. In such flows, perturbations can
grow only transiently during finite times
\cite{Balbus_Hawley92,Chagelishvili_etal97,Dimitrov_etal11}, which
is thought to be a key factor for the onset of subcritical
turbulence \cite{Grossmann00,Chapman02,Eckhardt_etal07}. One of the
basic characteristics of subcritical transition is its sensitivity
to the initial perturbation amplitude (e.g., Refs.~\cite{Baggett_etal95,
Schmid_Henningson01, Lesur_Papaloizou10, Duguet_etal10}), which is
also observed here. We found that there exists a critical amplitude
for initial velocity and magnetic field perturbations (at a given
$L_x$, ${\rm Re}$ and ${\rm Rm}$) below which turbulence is absent
-- there is only transient amplification insufficient to trigger
transition, which eventually decays due to dissipation. By contrast,
for initial amplitudes larger than the critical value a turbulent
transition does occur after a phase of large enough transient
growth, as is also evident from Fig. 1. Specifically, at ${\rm Re}={\rm
Rm}=5$ adopted here, the critical amplitude turned out to be
$\langle {\bf u}^2 \rangle^{1/2}_{crit}=\langle {\bf b}^2
\rangle^{1/2}_{crit}=0.34$ (for the same type of initial noise
spectrum for both velocity and magnetic field perturbations), and in
the fiducial run we accordingly selected the initial rms amplitudes
(=0.84) larger than this in order to achieve turbulent regime. This
confirms that the turbulence we study here is subcritical, however,
we have not explored the transition process, that is, have not
pinned down the critical transition amplitude for different values of
the system parameters (domain size, Reynolds numbers, etc.) in more
detail. The problem of subcritical transition in MHD shear flows
deserves a special investigation in its own right, but in the
present analysis we are mainly interested in the properties of the
resulting self-sustaining turbulence itself once it has settled into
quasi-steady state. The underlying physics of the onset and
sustenance of subcritical turbulence in spectrally stable HD shear
flows -- the bypass concept -- has been extensively studied
previously in a number of papers (see e.g.,
Refs.~\cite{Grossmann00,Eckhardt_etal07} for a review), but
extension to MHD turbulence in spectrally stable magnetized shear
flows, to the best of our knowledge, has not been systematically
investigated yet. The equilibrium flow considered here with a linear
spanwise shear of mean velocity and streamwise magnetic field is the
simplest but important example of such spectrally stable magnetized
shear flows that allows us to grasp specific processes determining the
onset, self-sustenance and spectral characteristics of MHD
turbulence in this kind of flow. Deeper insight into the dynamics
of such subcritical MHD turbulence can be gained by performing
an analysis in spectral space.

\begin{figure}[t]
\includegraphics[width=\columnwidth]{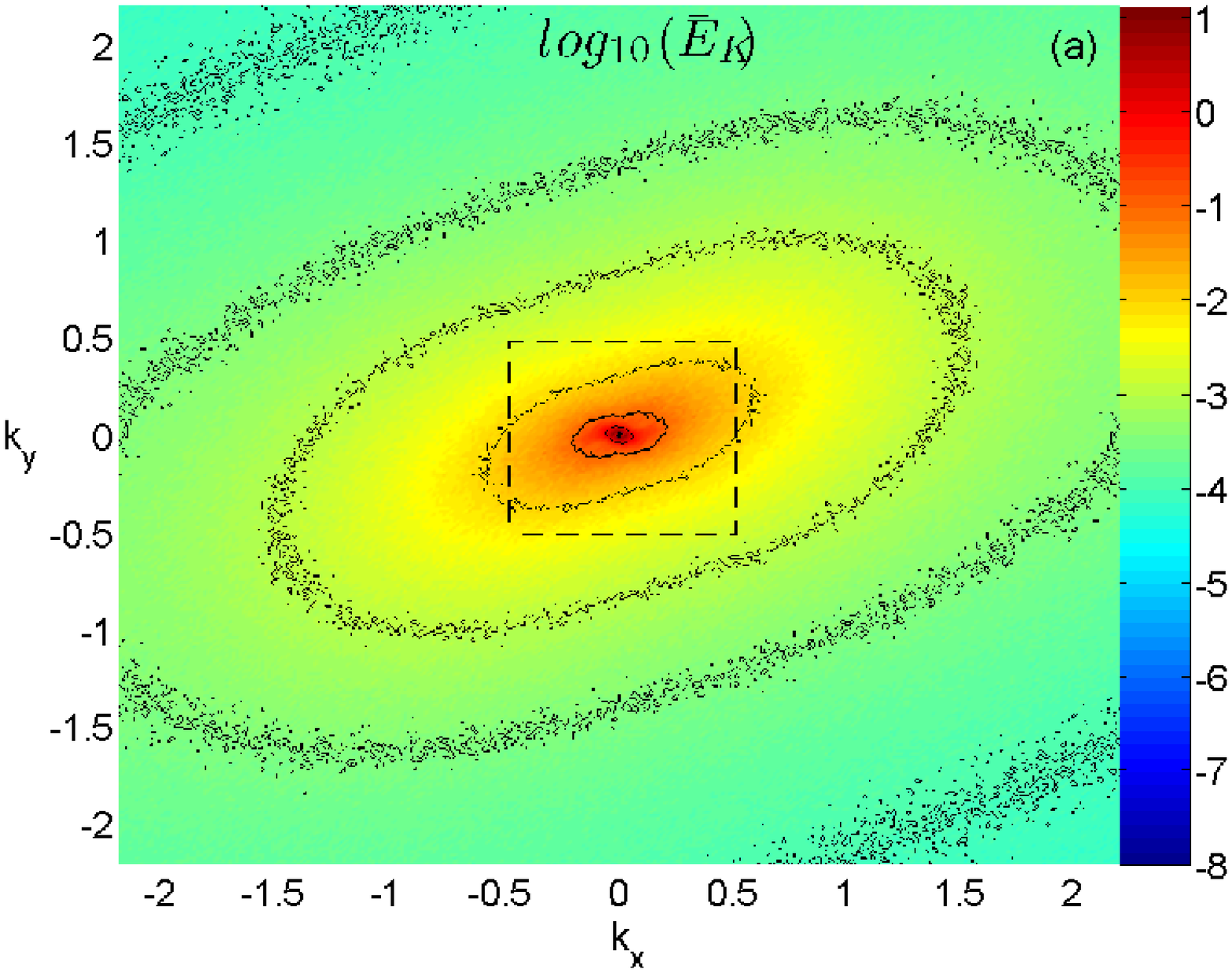}
\includegraphics[width=\columnwidth]{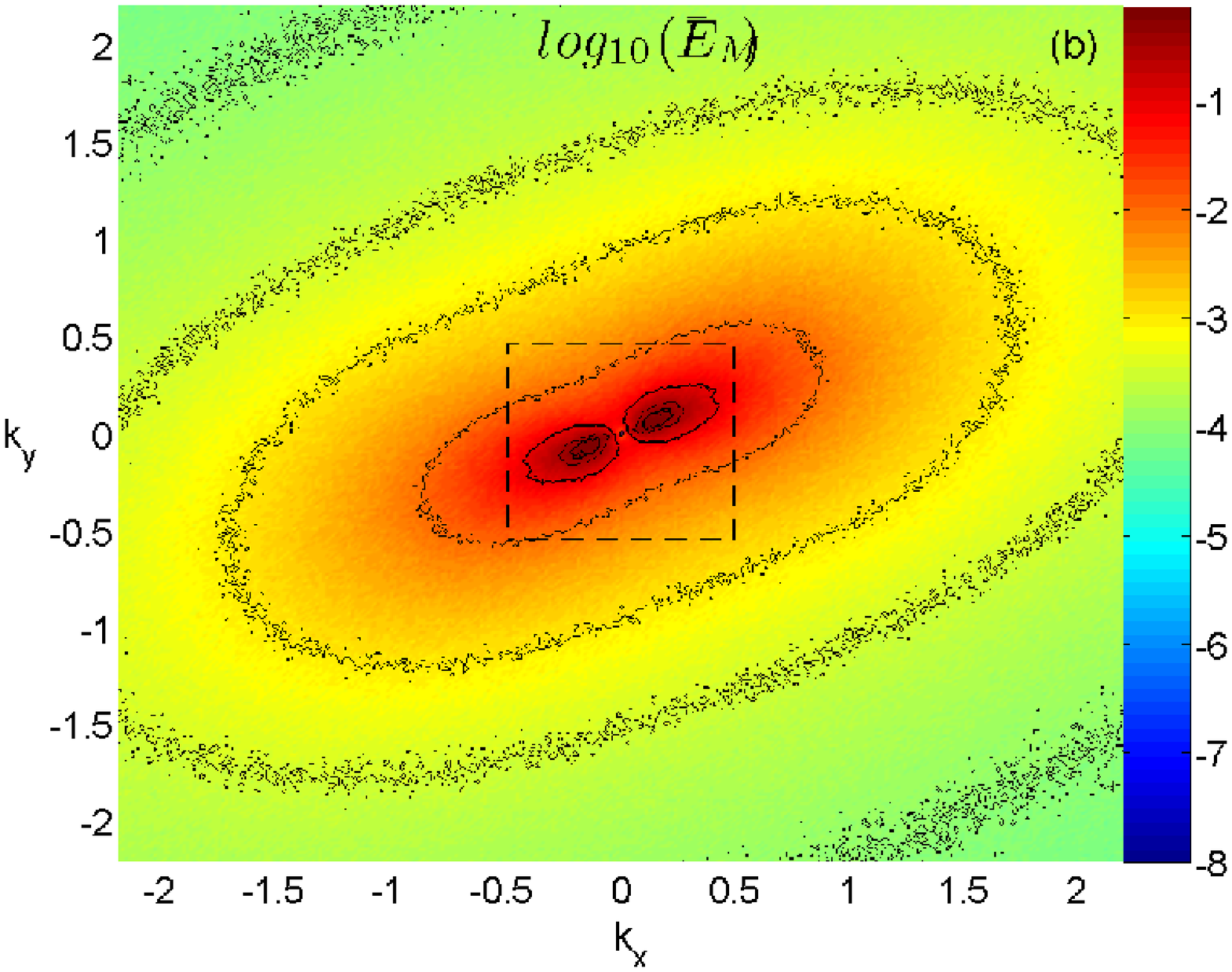}
\caption{(Color online) Time-averaged (a) kinetic and (b) magnetic
energies' spectra in ${\bf k}-$plane pertaining to the quasi-steady
turbulent state. These time-averages are done over 80 shear times,
as described in the text. The isolines correspond to
the values $-4, -3.5, -3, -2, -1, 0$ of $log_{10}(\bar{E}_K)$ in panel (a) and to
the values $-4, -3.5, -3, -2, -1, -0.5, -0.3, 0$ of
$log_{10}(\bar{E}_M)$ in panel (b). Both spectra are anisotropic,
having larger power at the $k_x/k_y>0$ side. The kinetic energy
spectra is more concentrated at smaller wavenumbers than the
magnetic one. The dashed rectangle in each plot encloses the region of
major activity of the dynamical terms in Eqs. (21) and (22), which
are shown in Fig. 5.}
\end{figure}

\section{Turbulence behavior in spectral plane}

In this section, we focus on the analysis of the dynamics of the
quasi-steady turbulent state in Fourier plane. We now explicitly
calculate the individual terms in Eqs. (21) and (22), which were classified and
described in Sec. II, using the simulation data. The
\textsc{snoopy} code, being of the spectral type, is particularly useful
for this purpose, as it allows us to directly extract Fourier
transforms from the data.

Before proceeding to spectral analysis, we note that generally a
turbulent field and hence its Fourier transform are quite noisy. To
remove this noise and extract valuable information on the trends in
the turbulence dynamics, all Fourier transforms (spectra) presented
below are averaged over 80 shear times. The interval between two
successive dumps in the code was set to $1$ shear time, so the
averaging is represented by 80 snapshots. From now on we concentrate
on the evolution after the quasi-steady saturated nonlinear state
has set in (i.e., at $t\gtrsim 250$), so we can choose the starting
moment for averaging arbitrarily over the duration of this state,
since the result is practically independent of this moment by virtue of
the quasi-steadiness of the process.

\subsection{Energy spectra}

Figure 3 shows the time-averaged spectra of the kinetic and magnetic
energies in ${\bf k}-$plane that have been established in the
quasi-steady turbulent state. Note that both spectra are strongly
anisotropic, with the magnetic energy spectrum being broader than
the kinetic energy one. For $k\gtrsim 0.5$, they have a similar
elliptical shape inclined to the $k_x-$axis, whereas at $k \lesssim
0.5$ these spectra differ in structure: isolines for the magnetic
energy divide into two sets of ellipses near the center with the same
inclination. This indicates that SFHs with $k_x/k_y>0$ have more
energy than those with $k_x/k_y<0$ at fixed $k_y$. Since $\beta \gg
1$, the effect of the mean flow shear prevails over that of the mean
magnetic field that leads us to suppose that the anisotropy of these
spectra might be primarily due to shear. \footnote{Similar
anisotropic spectra were also observed in the simulations of MHD
turbulence driven by the MRI in the presence of shear
\cite{Hawley_etal95,Lesur_Longaretti11}.} These features of the
kinetic and magnetic energy spectra, which clearly distinguish them
from typical turbulent spectra in the classical shearless case
\cite{Biskamp03}, arise as a consequence of the specific way in which the terms
of linear and nonlinear origin in Eqs. (21) and (22) operate in
${\bf k}-$plane. We show below that these terms are anisotropic
over wavenumbers due to shear, resulting in a new phenomenon -- the
transverse cascade of power in spectral plane -- compared to the
classical (isotropic) case.

The above time-averaged 2D spectra integrated over the angle in ${\bf
k}-$plane, $\bar{E}^{(k)}_{K,M}=k\int_0^{2\pi}\bar{E}_{K,M}d\phi$,
and represented as a function of $k$ are shown in Fig. 4. From
intermediate wavenumbers $k\sim 0.2$ up to dissipation wavenumbers
$k\sim k_D=2.24$, both one-dimensional (1D) spectra exhibit power-law
dependence on $k$, however, with different spectral indices -- the
kinetic energy spectrum is well fitted by $k^{-1.4}$ and the
magnetic energy spectrum by $k^{-2}$. At these wavenumbers, the
spectral density of the magnetic energy is larger than that of the
kinetic one, but at smaller $k\lesssim 0.2$ it decreases and becomes
less than the kinetic one, both deviating from the power-law. These
power-law parts of the spectra clearly differ from the typical
Iroshnikov-Kraichnan (IK) spectrum, $k^{-1.5}$, characteristic of
classical 2D and 3D MHD turbulence without background shear flow
\cite{Biskamp03}, though the kinetic energy spectrum is still close
to it. Different spectra of kinetic and magnetic energies, following
approximately power-laws (though, with kinetic energy spectrum
somewhat coincident with the IK one), are also present in analogous
3D simulations of MRI-driven MHD turbulence in the shearing box
model of a disk \cite{Simon_etal09,Fromang10,Lesur_Longaretti11}.
However, it was pointed out in those studies that in the presence of
differential rotation (shear) and weak magnetization ($\beta \gg 1$)
associated with disk flows, which are in fact also shared by the 2D
MHD shear flow considered here, classical Kolmogorov or IK
phenomenology is generally \emph{not} applicable to turbulence
dynamics, because due to shear, energy injection from the mean flow
into turbulence can occur over a broad range of length-scales
available in the flow, from the largest scale down to the dissipation
scale, that in turn prevents the development of the proper inertial
range of a spectrum in the classical sense (see also
Refs.~\cite{Fromang_Papaloizou07,Lesur_Longaretti11}). So, the
spectra obtained in those disk simulations, despite being of the
power-law type, are in fact determined by interplay between
injection terms due to the linear MRI, operating over a range of
wavenumbers, and nonlinear terms in spectral space. The situation is
similar in the present problem. As shown below, the action
of the energy injection terms $I_K$ and, especially, of $I_M$ extends
over a range of wavenumbers in ${\bf k}-$plane and is remarkably
anisotropic [see Figs. 5(a), 5(b) and 6]. As noted above, these
terms are responsible for the linear transient amplification of SFHs
and energy extraction from the mean flow, so in this respect they
play a similar role of supplying turbulence with energy in our
nonrotating case as the (transient) azimuthal MRI in rotating disk flows.
Moreover, we demonstrate that there exists a new phenomenon --
the transverse nonlinear cascade of spectral energy density --
resulting from this anisotropy and, ultimately, from shear. These new
features are not common to shearless MHD turbulence and hence it is
not surprising that Kolmogorov or IK theory cannot adequately
describe shear flow turbulence.

\begin{figure}[t]
\includegraphics[width=\columnwidth]{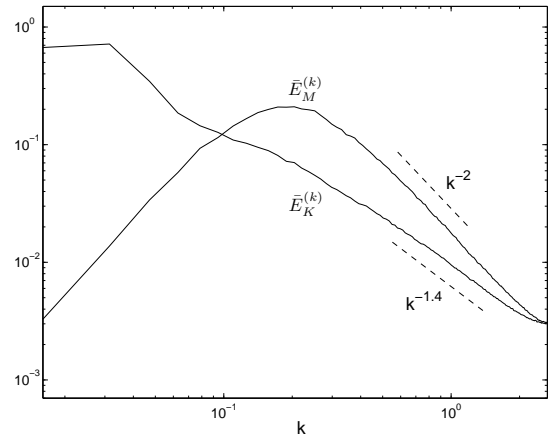}
\caption{Kinetic and magnetic energy spectra from Fig. 3 integrated
over the angle in ${\bf k}-$plane and plotted as a function of $k$. From
intermediate, $k\sim 0.2$, to dissipation, $k\sim k_D=2.24$,
wavenumbers a power-law behavior is observed in both spectra, though
with different spectral indices: $k^{-1.4}$ for the kinetic and
$k^{-2}$ for the magnetic energies.}
\end{figure}

We have presented the energy spectra from two perspectives: fully in
${\bf k}-$plane in Fig. 3 and their angle-integrated (over shells of
constant $|{\bf k}|$) versions in Fig. 4, the former is obviously
more informative than the latter. We emphasize that
angle-integration of turbulent spectra and transfer functions when
they are anisotropic in wavenumber plane might lead to the loss of
essential information on the detailed nonlinear dynamics, so we take
a more general strategy of Ref.~\cite{Horton_etal10} and represent
energy spectra as well as injection and nonlinear transfer terms in
full in ${\bf k}-$plane, in contrast to previous related studies of
MHD turbulence in shear flows considering either such
angle-integrated or reduced 1D spectra (e.g.,
Refs.~\cite{Fromang_Papaloizou07,Simon_etal09,Davis_etal10,Lesur_Longaretti11}).
This allows us to obtain a complete dynamical picture and
understanding of the nature of subcritical MHD turbulence in the
presence of mean flow shear.

\begin{figure*}
\includegraphics[width=\columnwidth]{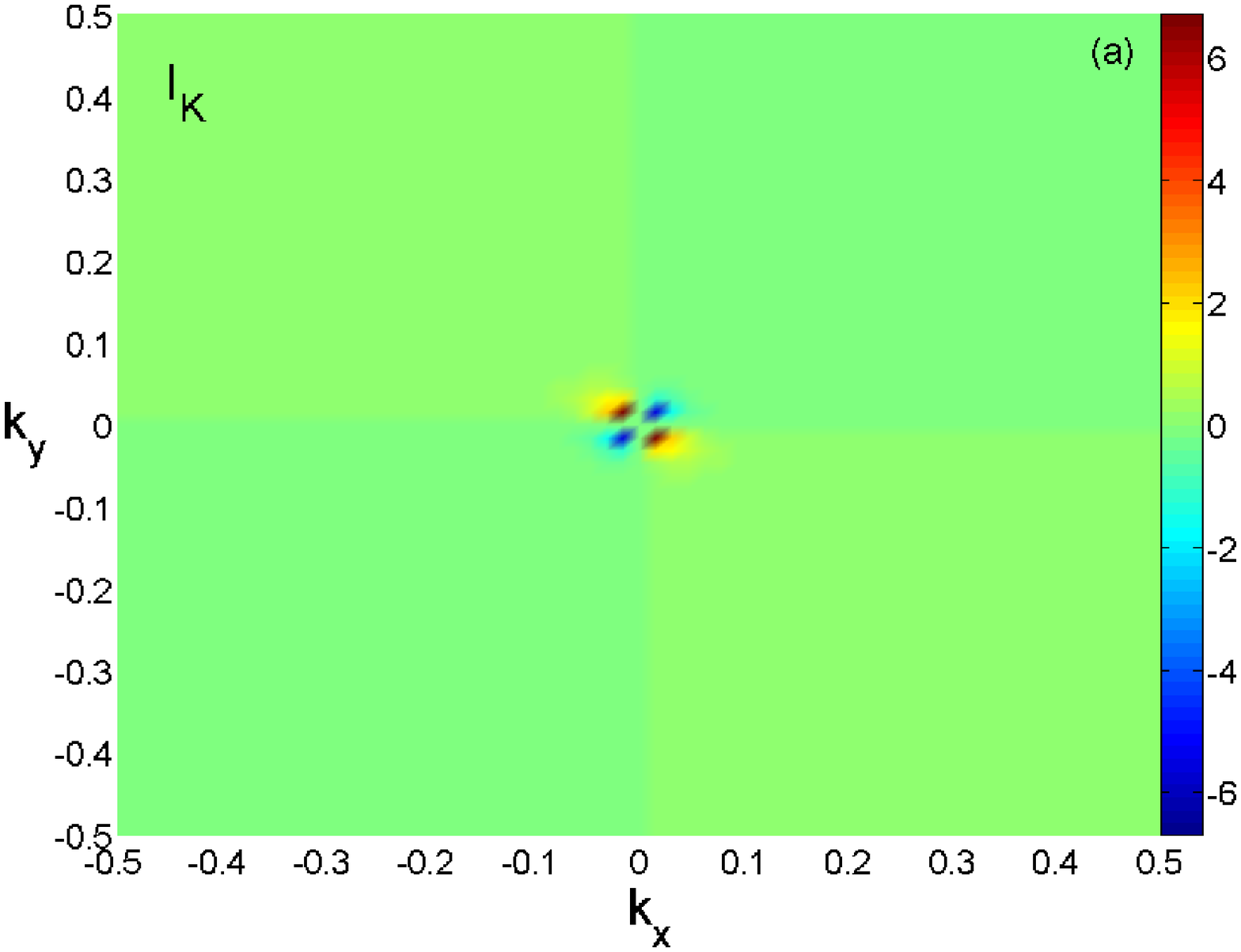}
\includegraphics[width=\columnwidth]{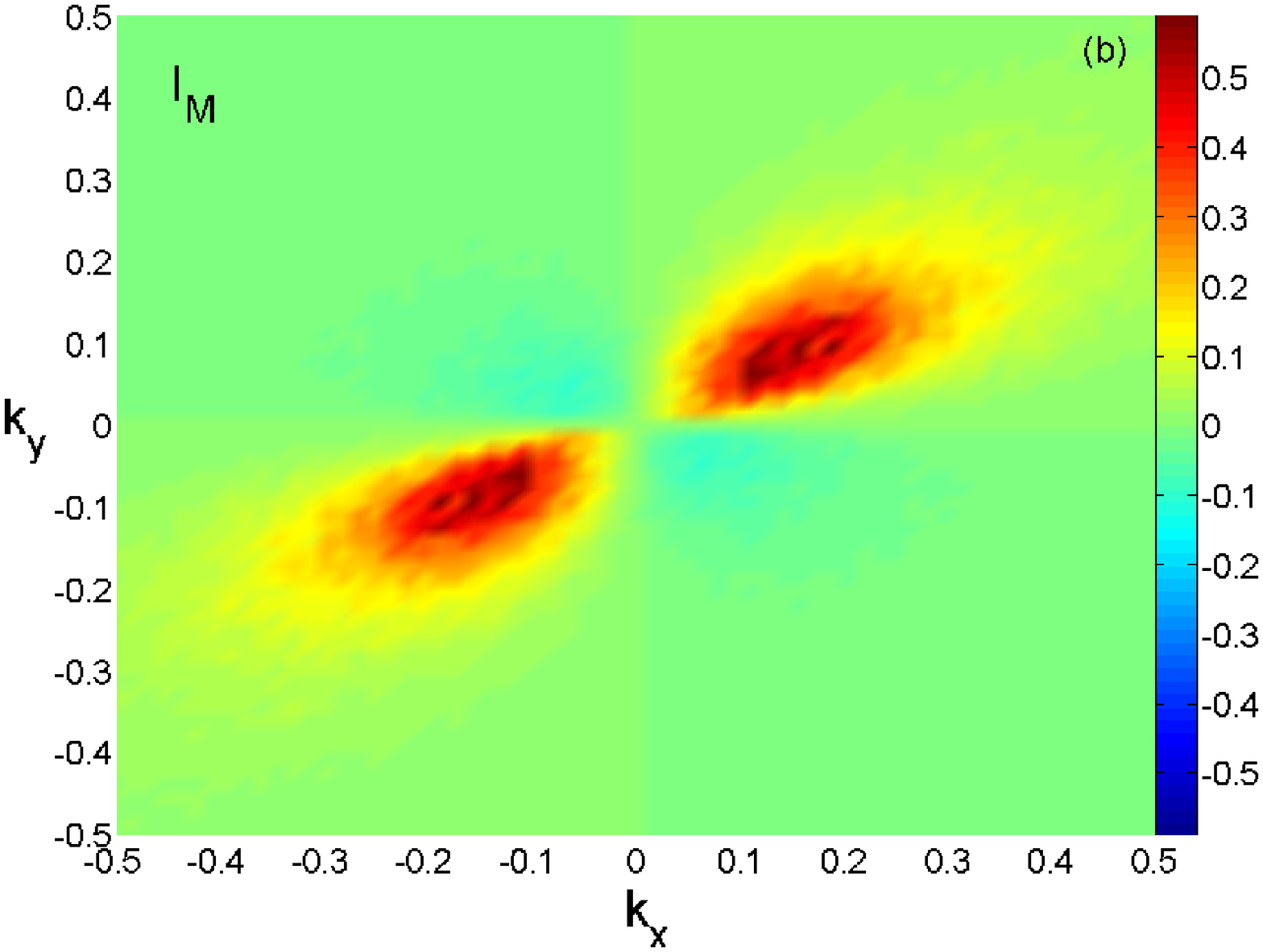}
\includegraphics[width=\columnwidth]{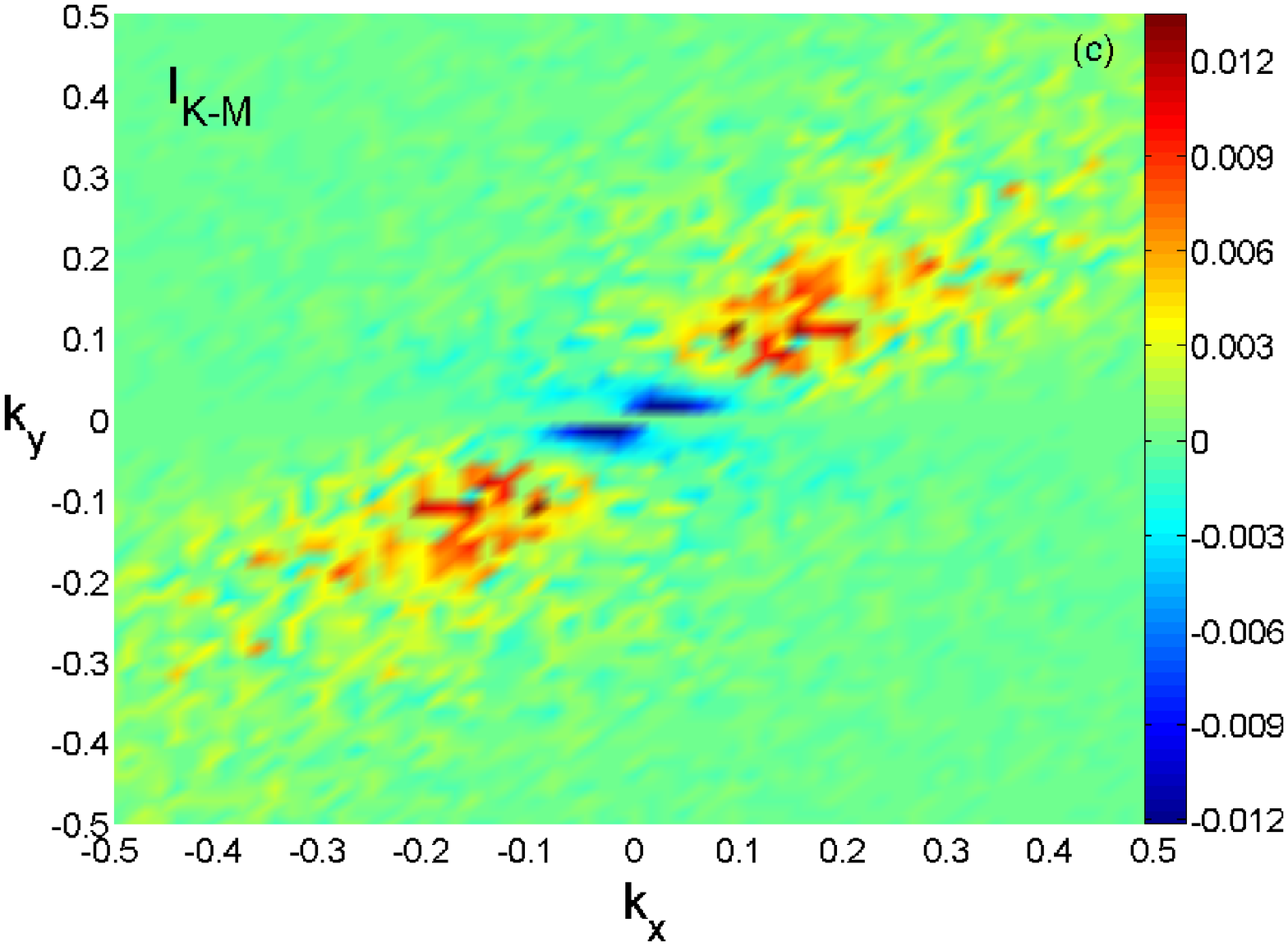}
\includegraphics[width=\columnwidth]{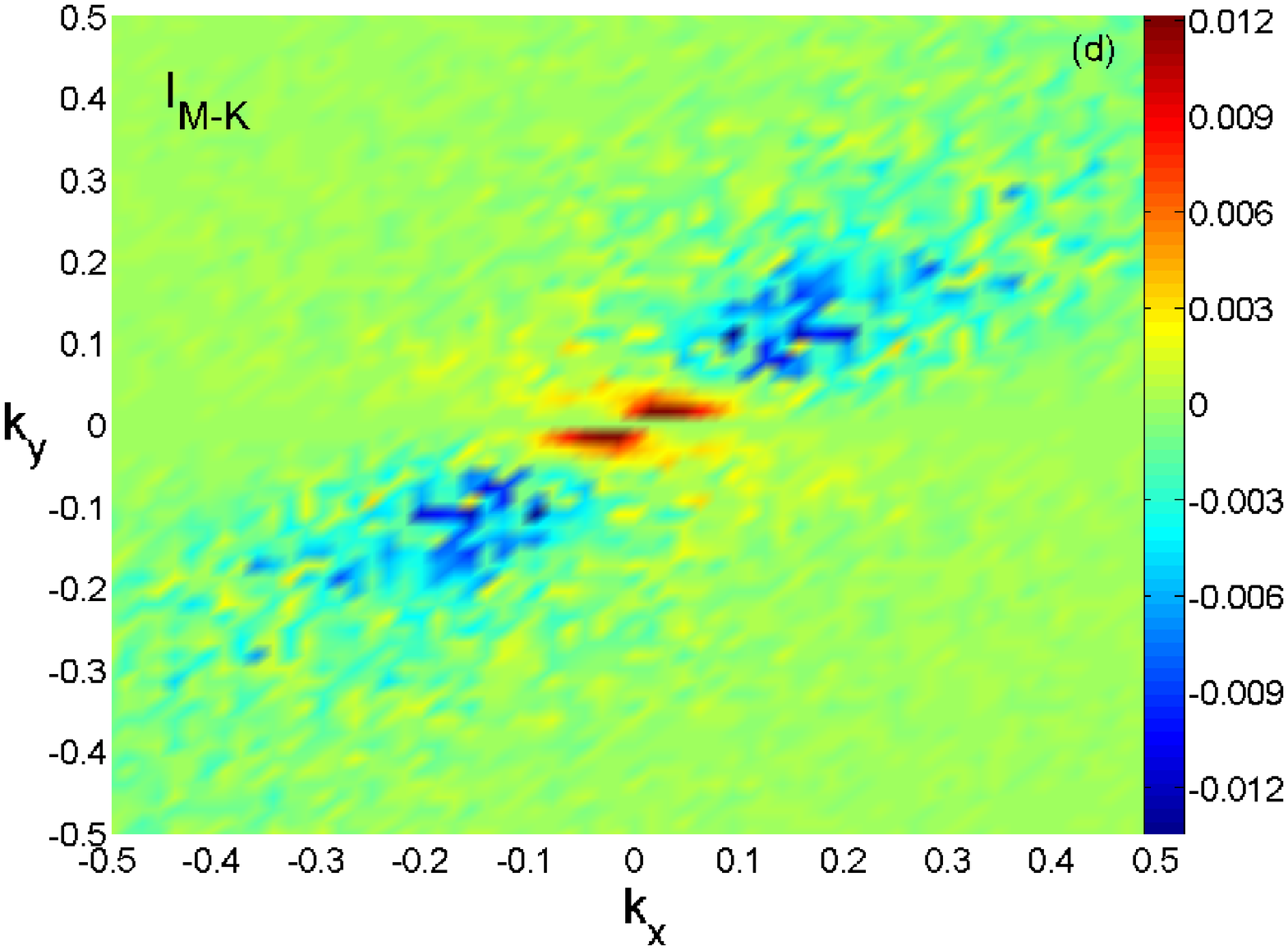}
\includegraphics[width=\columnwidth]{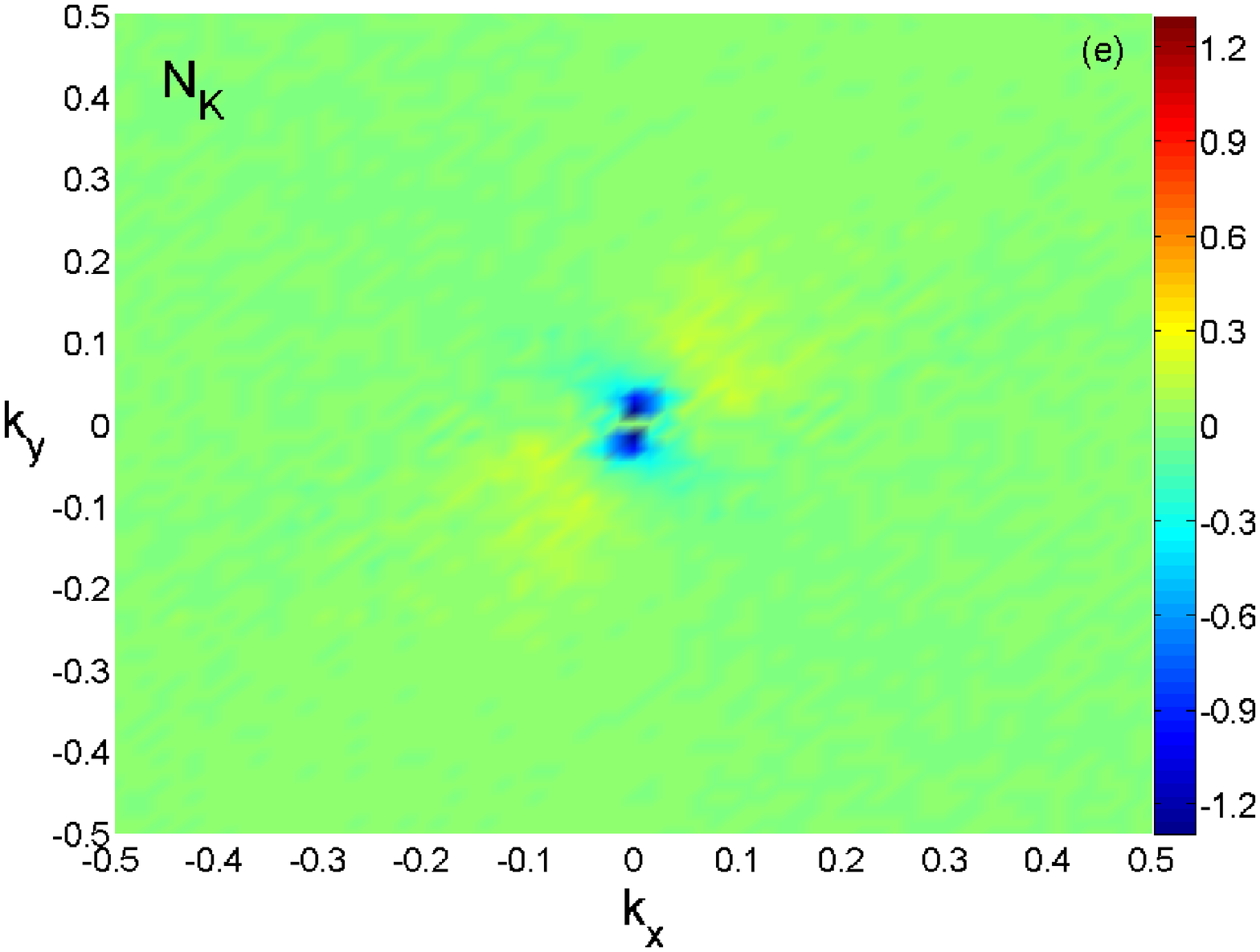}
\includegraphics[width=\columnwidth]{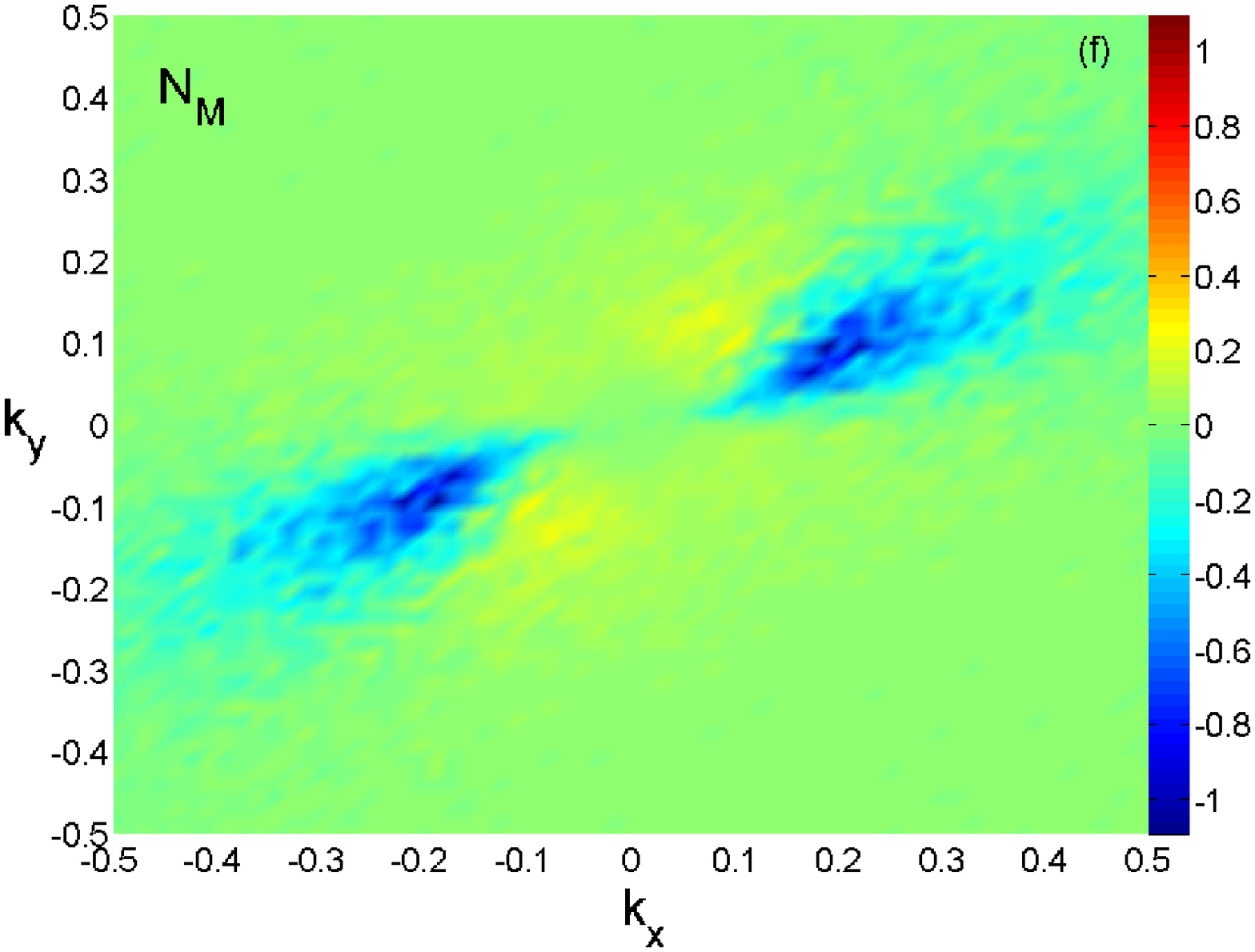}
\caption{(Color online) Maps of the time-averaged (a) kinetic,
$I_K$, and (b) magnetic, $I_M$, energy injection terms, [(c),(d)]
the cross terms $I_{K-M},I_{M-K}$ and the (e) kinetic, $N_K$, and
(f) magnetic, $N_M$, nonlinear transfer terms in ${\bf k}-$plane in
the state of quasi-steady turbulence. The time averages are obtained
over an interval of 80 shear times (from 472 to 552 shear times), as
described in the text. Kinetic energy injection mostly occurs at
small wavenumbers, $k\lesssim 0.1$, and on the $k_x/k_y<0$ side
where $I_K>0$, while magnetic energy injection occurs mostly at
intermediate wavenumbers, $0.05\lesssim k \lesssim 0.5$, on the
$k_x/k_y>0$ side where $I_M>0$, overall it is dominant over $I_K$,
i.e, energy injection into turbulence appears to be due mainly to
the Maxwell stresses. The $N_K$ and $N_M$ terms transfer,
respectively, the spectral kinetic and magnetic energies anisotropically
(transversely) in wavenumber plane, away from regions where they are
negative $N_K<0,~N_M<0$ (blue) to regions
where they are positive $N_K
>0,~N_M>0$ (yellow). The nonlinear terms are comparable to the injection terms and
both are about two orders of magnitude larger than the cross terms.}
\end{figure*}

\subsection{Spectra of energy injection: $I_K$ and $I_M$ }

To better understand the character of the above anisotropic kinetic
and magnetic energy spectra and nonlinear transfers, in Fig. 5 we
present the distribution of the time-averaged kinetic and magnetic
injection functions, $I_K$ and $I_M$, cross terms, $I_{K-M}$ and
$I_{M-K}$, and nonlinear transfer terms, $N_K$ and $N_M$, in
${\bf k}-$plane in the quasi-steady turbulent state. From this
figure it is seen that these terms differ in magnitude and, like the
spectral energies, all exhibit anisotropy over wavenumbers, that is,
depend on the wavevector angle. $I_K$ is mostly concentrated at
small wavenumbers, $k\lesssim 0.1$ [Fig. 5(a)], being positive at
$k_x/k_y<0$ (red and yellow regions), where it increases the
kinetic energy of SFH, and negative at $k_x/k_y>0$ (blue
regions), where it takes kinetic energy from SFH and gives it back to
the flow. A net contribution of $I_K$ over all wavenumbers  is,
however, negative (i.e., $\langle u_xu_y\rangle<0$). On the other
hand, $I_M$ mostly operates at larger wavenumbers, $0.05\lesssim
k\lesssim 0.5$ [Fig. 5(b)], and is dominant and positive on the
$k_x/k_y>0$ side (red and yellow regions), where it supplies
SFH with magnetic energy. The net result of $I_M$ over all
wavenumbers is a positive energy gain for perturbations (i.e.,
$\langle -b_xb_y\rangle>0$), which prevails over the net negative
effect of $I_K$, as is also evident from Fig. 1(b), and maintains
turbulence. So, energy input for perturbation SFHs is provided by
the magnetic source term $I_M$, which operates over a much broader region
in $\bf k$-plane than $I_K$ does. We checked that such a dependence
of kinetic and magnetic energy injection terms on wavenumbers, in
fact, is also seen for the linear evolution of SFH, i.e., when the
SFH drifts along the $k_x-$axis due to shear, its kinetic energy
first increases at $k_x/k_y<0$, then decreases after crossing the
point $k_x=0$, while its magnetic energy starts to increase at
$k_x/k_y>0$ during a few shear times and then continues to oscillate
with Alfv\'{e}n frequency, $\omega_A=u_Ak_y$, and constant amplitude
(provided dissipation is neglected).

The linear cross terms, $I_{K-M}$ and $I_{M-K}$ [Figs. 5(c) and
5(d)], are small compared to both $I_K,I_M$ and nonlinear $N_K,N_M$
terms. In spectral plane, the action of these terms is somewhat
opposite to that of the corresponding injection terms. $I_{K-M}$ lowers
the kinetic energy at small wavenumbers, but increases at intermediate
and large wavenumbers on the $k_x/k_y>0$ side, while $I_{M-K}$
lowers the magnetic energy at intermediate and large wavenumbers in the
same quadrant and increases it at small wavenumbers. As noted above,
these cross terms cancel out in the total energy Eq. (24) and
because they are much smaller than the other dynamical terms, do not
play any major role in the energy balance in Eqs. (21) and (22) too.

The difference between the injection wavenumbers for the kinetic and
magnetic energies is demonstrated more clearly in Fig. 6, showing these
injection, nonlinear transfer, and dissipation terms
angle-integrated in ${\bf k}-$plane, $I_K^{(k)},N_K^{(k)},D_K^{(k)}$
[Fig. 6(a)] and $I_M^{(k)},N_M^{(k)},D_M^{(k)}$ [Fig. 6(b)], and
represented as a function of $k$. It is seen from this figure that
the range of wavenumbers, where the injection terms are at work,
extends from the smallest wavenumbers in the domain, $k_{x,min}$, up
to $k\sim 1$, comparable to the dissipation wavenumber $k_D$.
$I_{K}^{(k)}$ is positive at small wavenumbers, reaching a maximum
at $k\approx 0.05$, then becomes negative and vanishing at $k>0.12$
(i.e., no longer injects kinetic energy). On the other hand,
$I_{M}^{(k)}$ is positive and hence creates the turbulence's magnetic
energy at all wavenumbers, reaching a maximum at $k\approx 0.2$,
which is about twice as large as that of $I_{K}^{(k)}$. Note in Fig.
6 that these injection and nonlinear transfer terms $N_{K}^{(k)}$ and
$N_{M}^{(k)}$ widely overlap. This implies that in the presence of
shear, there is not a single injection scale in the flow, as is
usually assumed in classical turbulence theory, but instead energy
injection occurs all the way from the largest length-scales down to
the dissipation scale. Therefore, although power-law spectra for both
the kinetic and the magnetic energies are found at $0.2\lesssim k
\lesssim 2$ (Fig. 4), they still cannot be considered as being a
proper inertial range, since energy is injected at these
intermediate scales (see also
Refs.~\cite{Fromang_Papaloizou07,Lesur_Longaretti11} for a similar
situation in the MRI-driven turbulence, where the injection of energy,
drawn from the mean flow, into turbulence occurs over a range of
scales at which nonlinear transfers operate as well). From Fig. 6,
it is also seen that in this wavenumber range, the dissipation terms
are much smaller than the injection and nonlinear transfer terms, so
this part of the energy spectra are in fact formed mainly as a
result of the combined action of the linear injection and nonlinear
cascade.

\begin{figure}[]
\includegraphics[width=\columnwidth]{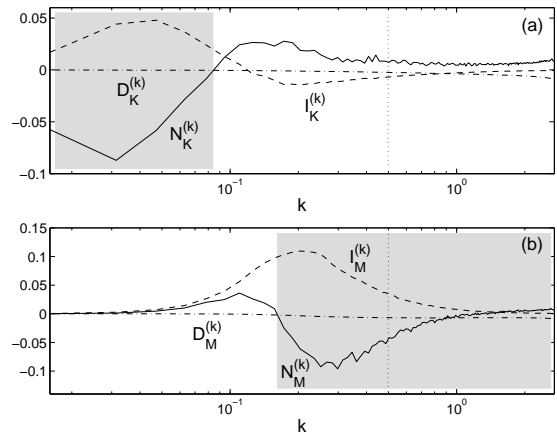}
\caption{(Color online) Kinetic and magnetic injection and nonlinear transfer terms
from Fig. 5 as well as dissipation terms integrated over the angle in
${\bf k}-$plane, (a) $I_{K}^{(k)},N_{K}^{(k)},D_{K}^{(k)}$ and (b)
$I_{M}^{(k)},N_{M}^{(k)},D_{M}^{(k)}$, and represented as a function
of $k$. Injection terms (dashed lines) operate over a range of
wavenumbers, overlapping with nonlinear terms (solid lines). The
magnetic energy injection is larger than the kinetic one. Both
viscous and resistive dissipation (dot-dashed lines) are relatively
important only at $k>k_D=2.24$. The reference dotted vertical line
marks the maximum wavenumber $k=0.5$ of the domains in Fig. 5. Shaded (gray)
regions correspond to wavenumbers at which $N_{K}^{(k)}<0,
N_{M}^{(k)}<0$ and hence the kinetic and magnetic energies,
respectively, are transferred, on average, away from these
wavenumbers due to nonlinearity.}
\end{figure}

\subsection{Nonlinear transfers $N_K$ and $N_M$ -- the essence of the transverse cascade}

We now move to describing the nonlinear kinetic and magnetic
transfer functions. As noted above, they do not represent a new
source of total energy for turbulence, but only act to redistribute
kinetic and magnetic spectral energies, which are extracted from the
mean flow, over wavenumbers and, in cooperation with injection
terms, determine the characteristics of spectra. So, our primary
goal is to understand how the nonlinear transfer terms work and,
consequently, in which directions energies cascade in Fourier plane
in the presence of background shear. As mentioned in Introduction,
for a purely HD constant shear (Couette) flow, which is spectrally
stable, it was shown in Ref.~\cite{Horton_etal10} that nonlinear
transfer function is anisotropic in ${\bf k}-$plane, i.e., depends
on the polar angle due to shear and, as a consequence, leads to
redistribution of the spectral energy over wavevector angles. This
relatively new process termed the angular, or transverse cascade of
energy has been shown to be essential for the maintenance of the
subcritical nonlinear state in this flow via the bypass mechanism.
Actually, identification of the transverse cascade of energy has
been made possible by virtue of representation of the dynamics fully
in 2D spectral plane, without performing angle-integration that
would result in washing out a key element of this process -- the
angular dependence (anisotropy) of the transfer functions' spectra. The
findings in that paper indicate that in HD shear flows, along with the
direct and inverse cascades quite well established in turbulence
theory, a new, transverse type of cascade can also take place which,
in fact, appears to be as important as the former. Based on these
results, in the present paper we generalize a spectral analysis of
nonlinear dynamics given in \cite{Horton_etal10} for the HD constant
shear flow to the MHD constant shear flow considered here, with the
aim of understanding the mechanism responsible for the sustenance of
the subcritical MHD turbulence in question. Specifically, we will
examine whether there exists a cooperative action of any kind between
energy-injecting linear and nonlinear transfer terms, like that
occurring in HD shear flows, capable of sustaining perturbations in
spectrally stable MHD shear flows.

Figures 5(e) and 5(f) show the distribution of the time-averaged
kinetic, $N_K$, and magnetic, $N_M$, nonlinear transfer functions
with wavenumbers in the quasi-steady turbulence, alongside the
injection terms, in order to easily see their cooperative
(correlated) action with the latter. As mentioned above, both $N_K$
and $N_M$ are strongly anisotropic, i.e., depend on the polar angle in
${\bf k}-$plane. This anisotropy has qualitatively the same
character as that of $I_K, I_M, I_{K-M}$ and the 2D energy spectra
in Fig. 3, that is, the spectra of all these are inclined towards
the $k_x-$axis due to shear. To bring out this angular
dependence more clearly, we integrated $I_K,I_M$ and $N_K,N_M$ over $k$, from the
smallest $k_{min}=k_{x,min}$ to the largest $k_{max}=k_{x,max}$
values in the domain,
\[
I_{K,M}^{(\theta)}=\int_{k_{min}}^{k_{max}}I_{K,M}kdk,~~
N_{K,M}^{(\theta)}=\int_{k_{min}}^{k_{max}}N_{K,M}kdk
\]
and represent them as functions of the polar angle $\theta$ in
Fig. 7. While the above-defined $N_K^{(k)}$ and $N_M^{(k)}$ describe
energy transfers in the direction of ${\bf k}$, $N_K^{(\theta)}$ and
$N_M^{(\theta)}$ describe energy transfer along the azimuthal
direction, perpendicular to ${\bf k}$.

\begin{figure}[t]
\includegraphics[width=\columnwidth]{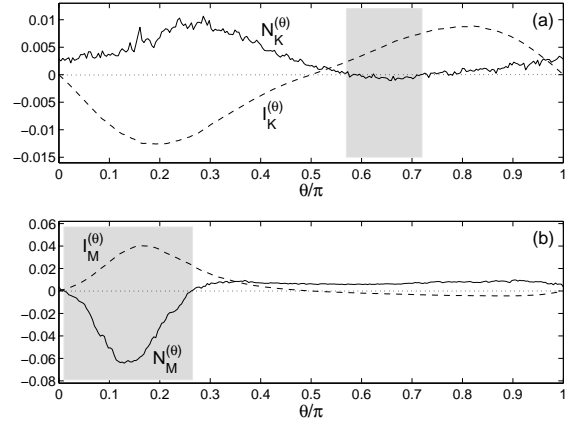}
\caption{(Color online) Kinetic and magnetic injection and nonlinear transfer terms
from Fig. 5 integrated over $k$, (a)
$I_{K}^{(\theta)},N_{K}^{(\theta)}$ and (b)
$I_{M}^{(\theta)},N_{M}^{(\theta)}$, and represented as a function
of the wavevector polar angle $\theta$ (angles $\pi<\theta<2\pi$
correspond to complex conjugates and are not shown here).
These plots clearly demonstrate the angular dependence (anisotropy)
of both the injection (dashed lines) and the nonlinear transfer (solid
lines) terms. Shaded (gray) regions correspond to angles at which
$N_{K}^{(\theta)}<0, N_{M}^{(\theta)}<0$ and hence kinetic and
magnetic energies, respectively, are transferred from these angles
to other angles due to nonlinearity, that is, a new phenomenon --
the transverse (angular) cascade of energy -- takes place.}
\end{figure}

As shown in Figs. 5-7, the distributions of $N_K$ and $N_M$ over
wavenumbers differ, leading to different types of cascades for the
kinetic and magnetic spectral energies. Since these quantities are
symmetric with respect to a change ${\bf k}\rightarrow -{\bf k}$,
without loss of generality, everywhere below we concentrate on the
upper part ($k_y>0$) of ${\bf k}-$plane. $N_K$ mainly operates in
two regions of ${\bf k}-$plane: at small wavenumbers, $k \lesssim
0.1$, where it is negative [blue region with
$N_K<0$ in Fig.5(e) corresponding to gray-shaded area with
$N_{K}^{(k)}<0$ in Fig. 6(a)], and at intermediate wavenumbers $0.1
\lesssim k \lesssim 0.5$ on the $k_x/k_y>0$ side ($0 \leq \theta
\leq \pi/2$), where it is positive (yellow
region with $N_K>0$); at all other wavenumbers the kinetic transfer
function is nearly zero. On the other hand, $N_M$ mainly operates at
$0.05 \lesssim k \lesssim 1$ [see also Fig. 6(b)], is positive at
$0.3\pi \lesssim \theta \leq \pi$ [yellow
region with $N_M>0$ in Fig. 5(f)] and negative at $0 \leq \theta
\lesssim 0.3\pi$ (blue region with $N_M<0$);
at all other wavenumbers the magnetic transfer term is nearly zero. Note
also that the distributions of $N_{K}$ and $N_{M}$ look somewhat
similar to those of the linear exchange terms $I_{K-M}$ and
$I_{M-K}$, respectively, but, as noted above, the latter are two
orders of magnitude smaller than the former.

By definition, these nonlinear transfer functions redistribute the
corresponding spectral energies away from the regions in ${\bf
k}-$plane where they are negative to the regions where they are
positive. The kinetic energy injection due to $I_K$ occurs, as
described above, at small wavenumbers ($k\lesssim 0.1$) with $\pi/2<
\theta < \pi$ where $I_K>0$ [see also Figs. 6(a) and 7(a)], but the
$N_K$ term is negative there, transferring kinetic energy away from
these injection wavenumbers to intermediate wavenumbers, $k\gtrsim
0.1$, with $0\leq \theta \leq \pi/2$, where $N_K>0$. This picture of
spectral kinetic energy transfer, or cascade towards larger
wavenumbers is also evident from Fig. 6(a), where the
angle-integrated $N_K^{(k)}$ changes from negative to positive at
about $k=0.1$, consistent with the flow of kinetic energy away from
$k\lesssim 0.1$ to $k\gtrsim 0.1$. The cascade behavior for the
turbulent magnetic energy is different from that of the kinetic
energy. The magnetic energy injection due to $I_M$ occurs at
intermediate wavenumbers ($0.05\lesssim k \lesssim 1$) for $0<
\theta < \pi/2$, where $I_M>0$ [see also Figs. 6(b) and 7(b)], but
the $N_M$ term, which is mostly negative there, transfers the
magnetic energy away from this injection region to its neighboring
region on the left with slightly smaller wavenumbers but larger
polar angles $0.3\pi \lesssim \theta \leq \pi$, where $N_M$ is
positive. This cascade of magnetic energy to smaller wavenumbers is
more clearly seen from Fig. 6(b), where the angle-integrated
$N_M^{(k)}$ changes from positive to negative at around $k=0.17$,
indicating the flow of magnetic spectral energy from $k \gtrsim 0.1$
to $k\lesssim 0.1$.

Thus, in shear MHD turbulence, the kinetic and magnetic energies are
transferred both along the wavevector, corresponding to familiar
direct and inverse cascades, and transversely (perpendicular) to it
(i.e., over angles $\theta$). Just this second type of nonlinear
cascade, better characterized by $N_K^{(\theta)}$ and
$N_M^{(\theta)}$ (Fig. 7), is a new effect of shear and is
discussed more in the next subsection; it is absent in classical
shearless MHD turbulence.

As stressed in Ref.~\cite{Horton_etal10}, the transverse cascade of
energy appears to be a generic feature of nonlinear dynamics of
perturbations in spectrally stable shear flows, so the conventional
description of shear flow turbulence solely in terms of direct and
inverse cascades, which leaves such nonlinear transverse cascade out
of consideration, might be incomplete and misleading. We emphasize
that in the present case revealing the complete picture of these
nonlinear cascade processes has become largely possible due to
carrying out the analysis in spectral plane. Because of the
shear-induced anisotropy of cascade directions, only
angle-integrated transfer functions in Fig. 6 (that are in fact
typically used in most numerical studies of shear MHD turbulence,
e.g., Refs.~\cite{Fromang_Papaloizou07,Simon_etal09,Davis_etal10,
Lesur_Longaretti11}), clearly, are not fully representative of the
actual, more general nonlinear redistribution of the spectral
energies in ${\bf k}-$plane, which also includes transfer with
respect to wavevector angles -- the transverse cascade.

\subsection{Interplay of the linear injection and nonlinear transverse cascade}

We have seen above that the nonlinear redistributions of spectral
kinetic and magnetic energies over the wavevector polar angle, $\theta$,
in ${\bf k}-$plane, termed the transverse cascade, are due to
shear-induced dependence of the nonlinear transfer functions $N_M$
and $N_M$ on this angle. This can be better appreciated from Fig. 7
showing the $N_K^{(\theta)}$ and $N_M^{(\theta)}$ introduced in previous
subsection. They exhibit different dependencies over $\theta$,
resulting in different characters of the transverse cascade for the
kinetic and magnetic energies. Note the opposite relative trends
between $I_K^{(\theta)}$ and $N_K^{(\theta)}$ [Fig. 7(a)] and
between $I_M^{(\theta)}$ and $N_M^{(\theta)}$ [Fig. 7(b)] with
respect to $\theta$:
\[
I_K^{(\theta)}\leq 0~~and~~N_K^{(\theta)}>0~~at~~0\leq \theta \leq
\pi/2,
\]
\[
I_K^{(\theta)}\geq 0~~and~~N_K^{(\theta)} \approx 0 ~~at~~\pi/2 <
\theta \leq \pi.
\]
On the other hand,
\[
I_M^{(\theta)}\geq 0~~and~~N_M^{(\theta)}\leq 0~~at~~0\leq \theta
\lesssim 0.3\pi,
\]
\[
I_M^{(\theta)}\approx 0~~and~~N_M^{(\theta)}> 0~~at~~0.3\pi \lesssim
\theta \leq \pi.
\]
This implies that the region of ${\bf k}-$plane, where SFHs are
replenished with kinetic energy by nonlinearity (i.e., where
$N_K>0$), lies on the right side of the kinetic energy injection
region with $I_K>0$, whereas the region where SFHs are replenished
with magnetic energy by nonlinearity (i.e., where $N_M>0$) lies on
the left side of the magnetic energy injection region with $I_M>0$,
as also seen in Fig. 5. As explained below, this specific
arrangement of the injection and nonlinear redistribution areas for
the magnetic energy in spectral plane appears to be crucial to the
sustenance of the turbulence.

After characterizing the specific activity of the linear injection
and nonlinear transfer terms in ${\bf k}-$plane associated with the
presence of shear, we now consider the evolution of SFHs in the
quasi-steady turbulence and identify a mechanism sustaining this
state. As noted above, apart from these terms, Eqs. (21) and (22)
also contain terms describing drift of SFHs in spectral plane due to
shear flow. In the upper half-plane ($k_y>0$) we focus on, all SFHs
drift along the $k_x-$axis direction and cross the injection and
transfer regions in succession. Since the turbulence is
quasi-steady, these three basic processes involved in the spectral
Eqs. (21) and (22): linear drift of SFHs, energy injection and
nonlinear transfer, together with viscous and resistive dissipation,
are in subtle balance, or cooperation, resulting in the closed
(positive) feedback loop that energetically maintains this state. We
interpret the workings of this loop as follows. Let us start the
loop cycle. The nonlinear transfer functions $N_K$ and $N_M$ supply
(from a previous cycle) SFHs with kinetic energy mainly at
wavenumbers with polar angles $0\leq \theta \lesssim 0.6\pi$ and
$0.7\pi \lesssim \theta \leq \pi$, where $N_K>0,N_K^{(\theta)}>0$,
and magnetic energy at $0.3\pi \lesssim \theta \leq \pi$, where
$N_M,N_M^{(\theta)}>0$ [see Figs. 5(e), 5(f) and 7]. Then, these
SFHs drift along the $k_x-$direction and enter the injection
regions, where $I_K>0$ and $I_M>0$. As a result, the kinetic energy
of those SFHs with $k_y\lesssim 0.1$ and the magnetic energy of
those SFHs with $k_y\gtrsim 0.05$ grow at the expense of the mean flow
-- just at this stage the kinetic and magnetic energies are being
injected into the turbulence due to $I_K$ and $I_M$ from the mean
flow. Then, the SFHs move into the regions where $N_K<0$ and $N_M<0$
and hence these nonlinear terms now act to transfer part of the
kinetic and magnetic energies from the amplified SFHs back,
respectively, to the regions where $N_K>0$ and $N_M>0$, from which
these SFHs started off, in this way regenerating new SFHs there
(positive nonlinear feedback). Towards the end of the cycle, part of
the original SFH's kinetic energy is returned to the mean flow,
since $I_K\leq 0$ at $0\leq \theta \leq \pi/2$, so effectively there
is no net gain of the turbulent kinetic energy from the mean shear
flow; the second part, which goes into the new SFHs, is taken from
the magnetic energy via the nonlinear exchange by positive $N_K$ (at
$k_y \gtrsim 0.1$) and the third part is gradually dissipated due to
viscosity as the SFH drifts further towards larger wavenumbers
($k\gtrsim k_D$). So, during each cycle, the SFHs gain primarily the
magnetic energy from the mean flow due to the injection term $I_M$.
Part of this magnetic energy is transformed by nonlinearity into the
kinetic, as mentioned above, and the other part into magnetic
energies of the newly created SFHs. The rest of the magnetic energy
is dissipated due to resistivity. As seen from Figs. 5(f) and 7(b),
in ${\bf k}-$plane, the magnetic injection region lies on the right
side of the region of its nonlinear regeneration where $N_M>0$. As a
consequence, these new (regenerated) SFHs will drift through the
same cycle and the whole process of (magnetic) energy extraction
from the mean flow will be repeated. In this way, a positive
feedback loop -- a cooperative interplay of the linear transient
amplification and nonlinear transverse redistribution of the
magnetic spectral energy is established, ensuring the sustenance of a
quasi-steady turbulent state at the expense of the background flow
energy. Such a constructive regeneration of those SFHs due to
nonlinearity, that can extract shear flow energy during the linear
transient amplification process, is the basis for the sustenance of
subcritical turbulence in spectrally stable shear flows in the
framework of the bypass concept \cite{Grossmann00}.

We have seen that a principal role in the above-described MHD
self-sustaining mechanism is played by magnetic field perturbations
that actually feed turbulence -- SFHs, which are able to extract
energy from the shear flow by means of the Maxwell stresses (i.e.,
by $I_M$), are continuously repopulated by the nonlinear magnetic
transfer term. This nonlinear positive feedback for the magnetic
perturbations is probably related to the fact that the Maxwell
stress has the ``right'' positive sign to supply turbulence [Fig.
1(b)]. By contrast, the injection region for the kinetic energy in
${\bf k}-$plane lies to the left and below the main region of its
nonlinear regeneration [at $0\leq \theta \leq \pi/2$ where $N_K>0$,
see Figs. 5(e) and 7(a)]. As a result, the majority of new SFHs,
drifting along the $k_x-$axis, cannot cross the injection region and
thus continuously gain the kinetic energy from the flow; even the
small fraction of new SFHs that can cross this region eventually
returns the kinetic energy to the flow where $I_K<0$. In other words,
the nonlinear feedback for the kinetic energy does not operate in a
similar, constructive, manner as that for the magnetic energy. This
may be related to the Reynolds stress being negative [Fig. 1(b)] and
hence ineffective in feeding turbulence with kinetic energy. So, in
the 2D MHD shear turbulence considered here, unlike the Maxwell
stress, the Reynolds stress cannot provide the right sign for transport.

\section{discussion and summary}

In this paper, we have studied the characteristics and self-sustaining
mechanism of subcritical MHD turbulence in incompressible magnetized
spectrally stable shear flows via DNS using
the spectral code \textsc{snoopy}. We have examined how the background
shear flow interacts with the turbulent fluctuations of the
incompressible 2D MHD equations to produce a self-sustained
turbulence. The analysis of the turbulence dynamics was carried out
in Fourier plane. To keep the problem as manageable as possible and
at the same time not to omit key effects of shear on the dynamics of
turbulence, as the base flow we took the simplest but important case
of plane MHD Couette flow with linear shear and an imposed
background uniform, weak, magnetic field parallel to it. This flow
configuration is linearly stable (with decaying linear perturbations
at long times) according to classical (modal) stability theory and
hence the only cause of transition to turbulence can be a linear
transient amplification of (magnetic field) perturbations  due to
the nonnormality associated with shear at streamwise wavenumbers
$k_y<S/u_A$. Consequently, the considered 2D MHD turbulence is
subcritical by nature. To understand its sustaining mechanism, we
Fourier transformed basic MHD equations and derived evolution
equations for the perturbed kinetic and magnetic spectral energies
in wavenumber plane. In these spectral equations, using the
simulation results, we calculated individual terms, which are
divided into two types -- terms of linear and nonlinear origin. The
terms of linear origin -- the Maxwell and Reynolds stresses -- are
responsible for energy exchange between the turbulence and the mean
flow through transient amplification of perturbation harmonics due
to shear. However, as we have shown, only the positive Maxwell stress
appears to be a dominant (magnetic) energy injector for the
turbulence; it is much larger than the Reynolds stress, which has a
negative sign and therefore does not contribute to the turbulent kinetic
energy gain. Another linear term due to shear in these equations
makes the spectral energies drift in the spectral plane parallel to
the $k_x-$axis. The nonlinear terms, which do not directly draw the
mean flow energy, act to transversely redistribute this energy in
Fourier plane, continually repopulating perturbation harmonics that
can undergo transient growth. Thus, we have demonstrated that in
spectrally stable shear flows, the subcritical MHD turbulent state is
sustained by the interplay of linear and nonlinear processes -- the
first supplies energy for turbulence via shear-induced transient
growth mechanism of magnetic field perturbations (characterized by
the Maxwell stresses) and the second plays an important role of
providing a positive feedback that makes this transient growth
process recur over long times and compensate for high-$k$
dissipation due to viscosity and resistivity.

This picture is consistent with the well-known bypass scenario of
subcritical turbulence in spectrally stable shear flows
\cite{Grossmann00} and differs fundamentally from a usual
(supercritical) turbulence scenario, which is based on exponentially
growing perturbations in a system that permanently supply turbulent
energy and do not require nonlinear (positive) feedback for its
sustenance. Such a cooperative action of linear transient growth and
nonlinear transfer mechanisms relies on anisotropy of the energy
spectra, injection and nonlinear cascades in spectral plane (see
Fig. 5), which is ultimately attributable to the flow shear. This
shear-induced anisotropy, i.e., the dependence of spectra and
nonlinear transfers on polar angle in ${\bf k}-$plane, as we found
and analyzed here in the case of MHD flows, appears to be inherent
in shear flow turbulence; a similar anisotropy exists in HD shear
flows (see Ref.~\cite{Horton_etal10} for details). It differs from
the typical anisotropy of classical (shearless) MHD turbulence in the
presence of a (strong) background magnetic field (e.g.,
Ref.~\cite{Goldreich_Sridhar95}) and changes the classical view on
nonlinear cascade processes: traditionally, the net action of
nonlinear turbulent processes is interpreted as either a direct or an
inverse cascade (e.g., Ref.~\cite{Biskamp03}). Our analysis
demonstrates, however, that in MHD shear flows, like HD ones, the
dominant nonlinear process, resulting from the spectral anisotropy,
is in fact the redistribution of perturbation spatial Fourier
harmonics over the wavevector \emph{angles}. (Probably for this reason,
in our simulations with background shear we did not observe the typical
2D coherent magnetic structures that grow via merging due to inverse
cascade of magnetic helicity \cite{Biskamp_Welter89, Wu_Chang01}).
These anisotropic energy transfers in Fourier space have been termed
\emph{nonlinear transverse redistribution}, or the transverse
cascade. In the considered flow, the nonlinear transverse cascade
plays a vital role in the long-term sustenance of turbulence -- it
redistributes mainly magnetic spectral energy over different angles
in ${\bf k}-$plane such that to continually regenerate those
harmonics which, drifting in spectral plane, have the potential to
undergo transient growth, extracting energy from the mean flow. This
indicates that the transverse cascade of spectral (magnetic) energy
appears to be characteristic of MHD turbulence in shear flows, so
the conventional characterization of nonlinear MHD cascade processes
in the presence of the flow shear in terms of direct and inverse
cascades, which ignores the transverse cascade, should be generally
incomplete and misleading. Identification of this new -- transverse
-- type of nonlinear cascades and its role in the maintenance of
shear MHD turbulence represents one of our main results.

We showed that as a result of anisotropy of nonlinear transfers in
${\bf k}-$plane, kinetic and magnetic energy spectra are also highly
anisotropic (see Fig. 3). These spectra integrated over wavevector
angle exhibit power-law behavior for intermediate wavenumbers,
though with different spectral indices: $k^{-1.4}$ for the kinetic
and $k^{-2}$ for the magnetic energies. Despite this, the
angle-averaged spectra we found should not be regarded as truly
inertial ranges, because the stresses inject kinetic and magnetic
energies into turbulence over a broad range of wavenumbers -- from
the largest scales in the domain down to the shortest scales
comparable to dissipation scale -- well overlapping with the
nonlinear transfer terms (see Figs. 6 and 7). So, these spectra are
determined by the combined effect of linear injection and nonlinear
transfer terms. This is in contrast to the usual forced turbulence case,
where energy is injected (by external forcing) in a narrow
wavenumber band and subsequent development of spectra is due to
nonlinearity only (e.g.,
Refs.~\cite{Biskamp03,Douglas_etal08,Newton_Kim09}). As noted above,
the energy injection by the stresses occurs through the transient
amplification of perturbation Fourier harmonics due to shear,
implying that the shear plays an important dynamical role at large
and intermediate scales ($\gtrsim u_A/S$). However, the
angle-averaging of anisotropic spectra (and also of transfer
functions) in shear flows, as often done in similar cases, might
result in the loss of essential information about the spectral
characteristics of shear turbulence because of its angular
dependence too.

In the context of the spectral indices, it is interesting to point
out that in some regions of the Earth's magnetotail, a magnetic energy
spectrum with a slope close to that obtained here, $k^{-2}$, is
observed \cite{Zimbardo_etal10}. It is hard to attribute this
observational result to either the Kolmogorov or the IK spectra. This
may suggest the influence of shear flow on the dynamics of the
magnetotail turbulence and formation of its spectrum. The way we see
it, definite conclusions can be drawn by performing a numerical
analysis similar to that presented here for a specific 3D model
configuration of the magnetotail.

In this paper, we have considered 2D dynamics and a brief discussion of
3D MHD turbulence in magnetized shear flows is in order. According
to the classical view, there is a fundamental difference in the
nonlinear dynamics of 2D versus 3D HD processes: 3D ones are
characterized by a direct cascade of energy, while 2D ones by
inverse cascade. By contrast, in MHD, the nonlinear dynamics of 2D
and 3D processes are similar in the sense that cascade directions of
characteristic quantities (energy, helicity, etc.) are identical
(see e.g., Ref.~\cite{Biskamp03}). As for the transverse cascade
analyzed in this paper, it occurs in HD as well as in MHD shear
flows. It is well-known that in HD shear flows, 2D turbulence is not
maintained and dies out (without external forcing), i.e., inverse
cascade modified by transverse cascade is unable to sustain
turbulence (HD turbulence in shear flows is usually 3D). The present study demonstrates that, unlike HD shear
flows, self-sustained 2D turbulence can do exist in MHD shear flows
owing to the transverse cascade. Being dependent on the shear, the
transverse cascade is expected to occur and play an important role
in the dynamics of 3D MHD shear turbulence too. But further studies
should clarify, whether the nonlinear dynamics with the third
$z-$direction (perpendicular to the flow plane) represents just a
mere extension of the basic self-sustaining process described here
in 2D or introduces a qualitatively new contribution. In any case,
the transverse cascade will remain a vital ingredient in the
self-sustenance of turbulence in 3D too. Although our analysis is
limited to 2D, since these are the streamwise and shearwise
directions, it allows us to bring out a basic mechanism underlying
the self-sustenance (via interplay of linear transient amplification
and nonlinear transverse cascade processes) and properties of
subcritical MHD shear turbulence.

Finally, we would like to discuss the applicability and relevance of our
approach to the MRI-driven 3D MHD turbulence in astrophysical disks.
Like the MHD shear flow considered here, disk flows are also weakly
magnetized \cite{Armitage11,Balbus03} and hence dominated by
shear-induced (transient) effects. Analysis of the dynamics of
MRI-turbulence in spectral space is important in order to understand
its basic nonlinear cascade properties, which play a decisive role
in various related processes such as the dependence of turbulence
saturation amplitude (turbulent transport) on viscosity and
resistivity (in terms of the magnetic Prandtl number,
\cite{Longaretti_Lesur10, Lesur_Longaretti07, Fromang_etal07,
Lesur_Longaretti11}), effective turbulent dissipation
\cite{Fromang_Papaloizou07,Simon_etal09}, emergence of large-scale
coherent structures (zonal flows,
\cite{Fromang_Nelson05,Johansen_etal09, Simon_etal12}) and dynamo
action
\cite{Brandenburg_etal95,Lesur_Ogilvie08,Davis_etal10,Guan_Gammie11,Bodo_etal12},
etc. A spectral analysis of fully developed MRI-turbulence in
magnetized disks has been carried out in a number of studies
\cite{Fromang_Papaloizou07,Simon_etal09,Davis_etal10,Lesur_Longaretti11},
as mentioned throughout the text. In these papers, the individual
terms in the evolution equation for the kinetic and magnetic
spectral energies are examined in wavenumber space, as also done
here. However, the main focus of these studies was on the
dissipative properties of turbulence, which depend on wavenumber
magnitude $k$ only, so energy spectra, injection and nonlinear transfer
functions angle-averaged in ${\bf k}-$space were used to
infer injection wavenumbers and cascade directions as well as the
dissipation wavenumbers. Evidently, such angle-integrated spectral
quantities give energy cascade features (direct and inverse) only
along the ${\bf k}-$direction. But, since one of the main causes of
the MRI in disks is shear associated with their differential
rotation (see e.g., \cite{Balbus03}), one would expect the dynamics
of the resulting turbulence to be essentially anisotropic in ${\bf
k}-$space (see also Refs.~\cite{Hawley_etal95,Lesur_Longaretti11}),
involving nonlinear transverse cascades, similar to those described
here, to be at work. This transverse cascade, arising from the
angular dependence of nonlinear spectral transfer functions, is
elusive under angle-integration and therefore was missing in these
studies. To the best of our knowledge, a more complete spectral
analysis of MRI-driven turbulence dynamics in 3D Fourier space
has not been done yet.

Such a spectral analysis is especially relevant and important for
understanding the nature of MRI-turbulence in zero net magnetic flux
and azimuthal (toroidal) magnetic field configurations, where the
linear MRI is manifested as transiently growing non-axisymmetric
modes \cite{Balbus_Hawley92,Hawley_etal95}, that is, no exponential
instability exists in these cases and hence the onset of the MHD
turbulence should be subcritical. This subcritical MRI-turbulence in
disks is currently the subject of active research in the disk
community. Although its characteristics in the presence of an imposed
non-zero net azimuthal field was studied extensively (e.g.,
Refs.~\cite{Hawley_etal95, Guan_etal09, Simon_Hawley09}), the main focus was on the
effects of viscosity and resistivity on the saturation properties of
turbulence, so no clear-cut picture of its basic sustaining
mechanism was presented. For zero net flux case, it is thought that
some type of MHD dynamo action must be operative, which generates a
large-scale azimuthal field able to sustain the turbulence (e.g.,
Refs.~\cite{Brandenburg_etal95,
Lesur_Ogilvie08,Davis_etal10,Guan_Gammie11,Bodo_etal12}). The
considered here configuration with a parallel magnetic field is in
fact equivalent to disk flows with azimuthal background field in the
local shearing box model (which in addition includes rotation). So,
based on this analogy, we speculate that the sustenance mechanism of
subcritical MHD shear turbulence presented here can be realized in
disk flows too and be responsible for a long-lived MRI-turbulence in
them. To investigate this in more detail, one should generalize a
similar type of spectral analysis of turbulence dynamics in 3D
Fourier space in disk flows with non-zero net azimuthal magnetic
field in the shearing box approximation.

\begin{acknowledgments}
We would like to thank Dr. G. Lesur for helping to familiarize with
the specifics of the \textsc{snoopy} code and Dr. A. G. Tevzadze for
discussions on the physical aspects of the problem. GRM acknowledges
financial support from the Rustaveli National Science Foundation.
\end{acknowledgments}

\bibliography{biblio}

\end{document}